\documentclass[conference,compsoc]{IEEEtran}
\usepackage[linesnumbered,ruled,vlined]{algorithm2e}
\usepackage{bm}
\usepackage{amsthm,amssymb,amsmath}
\usepackage{amsfonts}
\usepackage{multirow}
\usepackage{float}
\usepackage{graphicx}
\usepackage{booktabs}

\newtheorem{assumption}{Assumption}
\newtheorem{proposition}{Proposition}
\newtheorem{theorem}{Theorem}
\newtheorem{corollary}{Corollary}
\newcommand{\methodname}{{\tt{RIM}}}
\newcommand{\pvcg}{{\tt{PVCG}}}

\ifCLASSOPTIONcompsoc
  \usepackage[nocompress]{cite}
\else
  \usepackage{cite}
\fi
\ifCLASSINFOpdf
\else
\fi
\begin{document}
\title{A VCG-based fair incentive mechanism for federated learning}

\author{\IEEEauthorblockN{Mingshu Cong}
\IEEEauthorblockA{The University of Hong Kong \\
Email: mscong@cs.hku.hk}
\and
\IEEEauthorblockN{Han Yu}
\IEEEauthorblockA{Nanyang Technological University\\
Email: han.yu@ntu.edu.sg}
\and
\IEEEauthorblockN{Xi Weng}
\IEEEauthorblockA{Peking University\\
Email: wengxi125@gsm.pku.edu.cn} 
\and
\IEEEauthorblockN{Jiabao Qu}
\IEEEauthorblockA{LogiOcean Technologies, Ltd\\
Email: qujiabao@logiocean.com}
\and
\IEEEauthorblockN{Yang Liu}
\IEEEauthorblockA{WeBank\\
Email: yangliu@webank.com}
\and
\IEEEauthorblockN{Siu Ming Yiu}
\IEEEauthorblockA{The University of Hong Kong\\
Email: smyiu@cs.hku.hk}
}


% compsoc conferences compared to traditional conferences), use this
%Telephone: (800) 555--1212, Fax: (888) 555--1212}


\maketitle

\begin{abstract}
The enduring value of the Vickrey–Clarke–Groves (VCG) mechanism has been highlighted due to its adoption by Facebook ad auctions. Our research delves into its utility in the collaborative virtual goods production (CVGP) game, which finds application in realms like federated learning and crowdsourcing, in which bidders take on the roles of suppliers rather than consumers. We introduce the \underline{P}rocurement-\underline{VCG} (\pvcg{}) sharing rule into existing VCG mechanisms such that they can handle capacity limits and the continuous strategy space characteristic of the reverse auction setting in CVGP games. Our main theoretical contribution provides mathematical proofs to show that \pvcg{} is the first in the CVGP game context to simultaneously achieve truthfulness, Pareto efficiency, individual rationality, and weak budget balance. These properties suggest the potential for Pareto-efficient production in the digital planned economy. Moreover, to compute the PVCG payments in a noisy economic environment, we propose the \underline{R}eport-\underline{I}nterpolation-\underline{M}aximization (\methodname{}) method. \methodname{} facilitates the learning of the optimal procurement level and \pvcg{} payments through iterative interactions with suppliers.
\end{abstract}


% \fi
\IEEEpeerreviewmaketitle

\section{Introduction}

Economics, rooted in Adam Smith's concept of the invisible hand, asserts that perfect competition leads to \emph{Pareto efficiency}, a state where no one can benefit without disadvantaging another. This principle has been seriously challenged in the digital age, as the production of virtual goods may not lead to a Pareto-efficient state. Virtual goods can be duplicated without any expense, and a second producer could \emph{free ride} by copying the output of the initial producer and thus bypassing associated production costs.

Classical economics asserts that a benevolent central coordinator is needed to restore efficiency in the presence of free riding. In the context of the digital age, Web services empowered by big data inherently lean toward a centralization of AI power. Hence, there is a looming possibility that, in the near future, a central coordinator can emerge if we can properly harness the AI power. If such a situation does occur, we want to explore how a central coordinator can best realize Pareto-efficient production of virtual goods. 

In essence:
\emph{The economy is transitioning to Web-driven circulation and AI-empowered central planning.}

\begin{figure*}
\includegraphics[width = \textwidth]{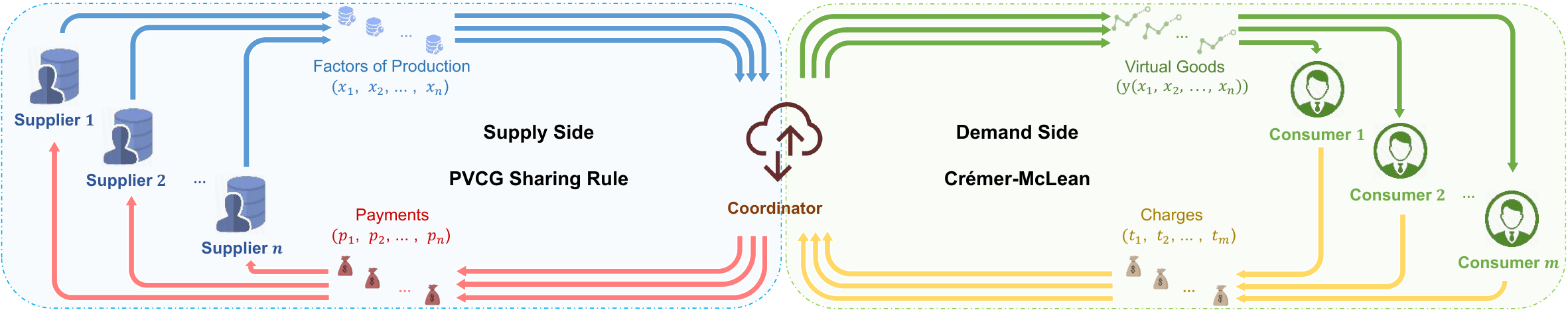}
\caption{Circular flow diagram of Web-driven collaborative virtual goods production and consumption.}
% \Description{Circular Flow Diagram of Collaborative Virtual Goods Production and Consumption in Digital Age.}
\label{fig:teaser}
\end{figure*}

\begin{figure}[!t]
  \centering
  \includegraphics[width =\linewidth]{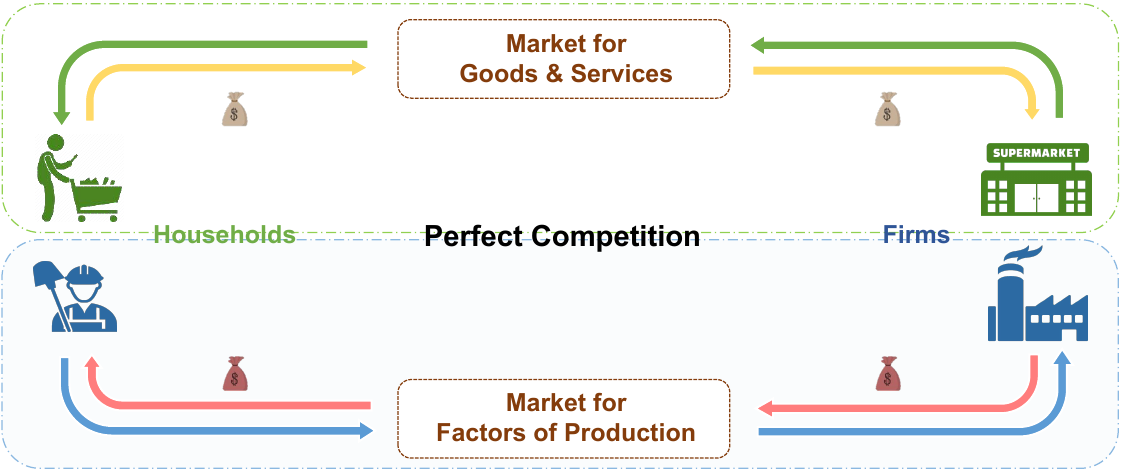}
  \caption{Circular flow diagram of classical economy.}
  % \Description{Circular Flow Diagram of Two-Sector Classical Economy.}
  \label{fig:circularClassical}
\end{figure}

However, incorporating a central coordinator into a perfectly competitive market economy is challenging, as it fundamentally alters the core tenets of classical economics. The circular flow diagram of a two-sector classical market economy, as depicted in Fig.\ref{fig:circularClassical}, has to evolve into Fig.\ref{fig:teaser} in the context of collaborative virtual goods production and consumption. Notably, the \emph{price discovery} process in a classical economy, facilitated by perfectly competitive markets, gives way to game-theoretic \emph{mechanisms} designed by the coordinator. This economic model is exemplified in applications like federated learning~\cite{yang2019federated} and crowdsourcing~\cite{crowd_howe2006rise}.

In this study, we will answer the following question:
\emph{Can we incentivize efficient production in this next era of digital planned economy?}

Here, our focus narrows to the supply-side \emph{mechanism design} illustrated in the left segment of Fig.\ref{fig:teaser}. The VCG mechanism\cite{vickrey1961counterspeculation}, adopted by Facebook ad auctions, has become increasingly significant in the computer science realm. We follow the idea of the VCG mechanism and propose a new mechanism called the \underline{P}rocurement-\underline{VCG} (\pvcg{}) sharing rule to overcome the free rider problem in virtual goods production. We introduce the interactive  \underline{R}eport-\underline{I}nterpolation-\underline{M}aximization (\methodname{}) method to implement the \pvcg{} sharing rule in a noisy economic environment. 
Basically, our \pvcg{} sharing rule differentiates itself from Facebook's classical VCG auction in three distinct manners: (1) \pvcg{} functions as a reverse auction where bidders act as suppliers for factors of production; (2) it assumes capacity limits for each supplier; and (3) its strategy space is continuous by design. This paper aims to elucidate how \pvcg{} ensures truthfulness. When combined with the demand-side Cr{\'e}mer-McLean mechanism~\cite{cremer-mclean-creemer1985optimal}, \pvcg{} reinstates the ideal of Pareto efficiency, hinting at the potential for efficient central planning in our digital age.

\subsection{Our Contribution}

We highlight the key contributions of our work as follows:
\begin{itemize}
    \item We are the first to adapt the VCG mechanism to a reverse auction setting with distinct features: capacity limits for suppliers and a continuous strategy space.
    \item \pvcg{} is the first mechanism tailored for the collaborative virtual goods production (CVGP) game that simultaneously ensures truthfulness, Pareto efficiency, individual rationality, and weak budget balance.
    \item We propose the \underline{R}eport-\underline{I}nterpolation-\underline{M}aximization (\methodname{}) method to learn the optimal procurement level and \pvcg{} payments through finite interactions with the suppliers.
\end{itemize}

\section{Related Work}

The Vickrey-Clarke-Groves (VCG) mechanism, proposed half a century ago by \cite{vickrey1961counterspeculation,groves1973incentives,clarke1971multipart}, has stood the test of time. Varian dedicated a chapter to elucidate the VCG mechanism in his renowned textbook on intermediate microeconomics \cite{intermediate-varian1995intermediate}, ensuring its recognition among economists. Facebook adopted the VCG mechanism in 2010 for its ad auctions. For a comprehensive understanding of the VCG mechanism's application to online auctions, ~\cite{facebook-varian2014vcg, facebook-profiting-leo2020vcg} serve as excellent resources.  Recent studies on the VCG mechanism can be found in works such as ~\cite{vcg_grubenmann2018financing, vcg_fukuta2011toward, vcg_mehta2022auction, vcg_ad_deng2023autobidding}.

Our study focuses on the supply-side issue where bidders act as suppliers for factors of production. This is commonly termed a \emph{reverse auction} \cite{reverse-jap2002online, revere-chen2012reverse} or a \emph{procurement game} \cite{chen2005efficient,iyengar2008optimal,drechsel2010computing}. A \emph{continuous game} involves a finite set of players with continuous strategy spaces\cite{continuous-ratliff2016characterization, continuous_mazumdar2020gradient}.  Existing literature has explored the VCG mechanism's role in such continuous reverse auctions, often employing the gradient descent method for payment determination
\cite{vcg-reverse-continuous-karaca2019designing, vcg-reverse-continous-angeli2023gradient}. 

Our CVGP game is suitable for domains like federated learning \cite{yang2019federated} and crowdsourcing \cite{crowd_howe2006rise}. Notably, the federated learning community has incorporated key objectives of mechanism design---including incentive compatibility, social maximization, individual rationality, and budget balance---into the IEEE Federated Machine Learning (P3652.1) Standard \cite{ieee_standard_yang2021white}. This integration has stimulated considerable research into the incentive mechanisms of federated learning \cite{fl_incentive_zhang2021incentive, fl_incentive_imteaj2021survey, fl_incentive_zeng2021comprehensive, fl_intentive_tu2022incentive, fl_incentive_ma2022applying, fl_incentive_shi2023towards, fl_incentive_ali2023systematic, fairness-yu2020fairness, framework-cong2020game}.

Though not the primary focus of our study, applying the Cr{\'e}mer-McLean mechanism \cite{cremer-mclean-creemer1985optimal} on the demand side allows us to separate the supply side from the demand side by fully extracting consumer surplus. Techniques for implementing this mechanism have been discussed in \cite{cremer-computation-albert2015assessing, cremer-computation-albert2017mechanism}.

\section{High-Level Idea}

\subsection{Continuous Second-Price Auction}

Recall that the VCG mechanism generalizes the concept of the second-price sealed-bid auction (a recap is provided in Appendix ~\ref{sec:vcgRecap}). Let $S$ represent the social surplus and $S_{-i}$ denotes the social surplus excluding bidder $i$. Then, the VCG payment $p_i$ to bidder $i$ can be expressed as:
\begin{equation}
    \label{eq:second-price}
    p_i = S_{-i}(\bm{x}^*) - S_{-i}(\bm{z}_{i}^*)  .
\end{equation}
Here,  $\bm{x}^*$ indicates the optimal allocation with all bidders included, and $\bm{z}_i^*$ signifies the optimal allocation excluding bidder $i$. Our discovery is that this expression is universally applicable to both discrete and continuous strategy spaces.

In the context of our CVGP game, the social surplus $S(\cdot)$ is:
\[
    S(\bm{x}) = r(\bm{x}) - \sum_{i=1}^n c_i(x_i),
\]
where $r(\bm{x})$ represents the coordinator's revenue and $c_i(x_i)$ indicates the cost function of the $i$th supplier. It is assumed here that the Cr{\'e}mer-McLean mechanism fully extracts consumer surplus.

Excluding supplier $i$ from the game is equivalent to setting its capacity limit to $0$. Thus, we express as:
\[
    \bm{x}^* = \mathop{\textrm{argmax}}\limits_{\bm{x} \le \bar{\bm{x}}} S(\bm{x}), \quad
    \bm{z}_i^* =\mathop{\textrm{argmax}}\limits_{\bm{z} \le (0, \bar{\bm{x}}_{-i})} S(\bm{z}).
\]
Then, the \pvcg{} payments can be computed by using  Eq.~\eqref{eq:second-price}.

\subsection{Learn the Optimal Procurement Level via Gradient Descent}

To determine the optimal $\bm{x}^*$ that maximizes $S(\bm{x})$, the coordinator employs the gradient descent method, requiring suppliers to report their marginal costs for varying procurement levels.

During the $t$th interactive epoch, the gradient of the social surplus is given as:
\[
\nabla S^{(t)} =   \nabla r(\bm{x}^{(t)}) -  (c'_1(x_1^{(t)}), \cdots, c'_n(x_n^{(t)})) .
\]
This is determined by the observed revenue gradient and reported marginal costs.

Leveraging the gradient descent method allows us to locate the optimal procurement level
 $\bm{x}^*$.

\subsection{Interpolation}

We interpolate the gradients at unseen procurement levels using the historical gradients from historical procurement levels.

For revenue interpolation, we use the following form of the revenue function:
\[
    r(\bm{x}) = \phi(\bm{w}^T \bm{x}).
\]

Assume we have obtained the interpolated curves of $\phi'(\cdot)$ 
and $c_i'(\cdot)$, represented as $\hat{\phi'}(\cdot)$ and $\hat{c'_i}(\cdot)$, respectively. 
Then, the coordinator can determine the vector $\hat{\bm{z}}_i^*$ that maximizes $\hat{S}_{-i}(\cdot)$ by using the interpolated gradients.

Furthermore, we can demonstrate that:
\begin{align*}
   \hat{p}_i 
   & = \hat{S}_{-i} (\bm{x}^*) - \hat{S}_{-i} (\bm{z}_i^*) 
\\
&
    =  - \int_{\bm{w}^T \bm{x}^*}^{\bm{w}^T \bm{z}_i^*} \hat{\phi'}(s) ds
    + \sum_{k=1,k\ne i}^n \int_{x_k^*}^{z_{ik}^*} \hat{c_k'}(s)ds .     
\end{align*}

As a result, we are able to determine the \pvcg{} payments by computing definite integrals of the interpolated derivative curves.

\section{Preliminaries}

\subsection{Notation}
We use the following notations throughout this paper:
\begin{itemize}
    \item $i$: Index of a supplier, starting at $1$.
    \item $n$: Total number of suppliers.
    \item $j$: Index of a consumer, starting at $1$.
    \item $m$: Total number of consumers.
    
    \item $t$: Count of the current interactive epoch.
    \item $x_i$: Procurement amount from the $i$th supplier.
    \item $\gamma_i$: Cost type of the $i$th supplier.
    \item $c_i(x_i) = c(x_i, \gamma_i)$: Cost incurred by the $i$th supplier for supplying $x_i$ units of resources. 
    \item $\bar{x}_i$: Capacity limit of the $i$th supplier. 
    \item $\hat{x}_i$: Reported capacity limit of the $i$th supplier.
    \item $\hat{\gamma}_i$: Reported cost type of the $i$th supplier.
    
    \item $r(\bm{x})$: Coordinator's revenue at the procurement level $\bm{x}$.
    \item $\theta_j$: Valuation type of the $j$th consumer.
    \item $v_j(\bm{x}) = v(\bm{x}, \theta_j)$: Valuation of the $j$th consumer on the virtual good produced from  $\bm{x}$.
    
    \item $S(\bm{x}, \bm{\gamma})$: Social surplus given cost types $\bm{\gamma}$.
    \item $S_{-i}(\bm{x}, \bm{\gamma})$: Total surplus excluding  supplier $i$.
    \item $\bm{x}^*$: Optimal procurement level with all suppliers included.
    \item $\bm{z}_i^*$: Optimal procurement level excluding supplier $i$.
    \item $\mathcal{S}(\hat{\bm{x}}, \hat{\bm{\gamma}})$: Maximum social surplus based on reported capacity limits $\hat{\bm{x}}$ and reported cost types $\hat{\bm{\gamma}}$.
    \item $\mathcal{S}_{-i}(\hat{\bm{x}}, \hat{\bm{\gamma}})$: Maximum total surplus excluding supplier $i$.
    \item $(0, \bar{\bm{x}}_{-i})$: Capacity limit vector with the $i$th element being $0$.
    
    \item $\hat{r}(\cdot)$: Interpolated revenue curve.
    \item $\hat{c}_i(\cdot)$: Interpolated cost curve for supplier $i$.
    \item $\hat{c_i'}(\cdot)$: Interpolated marginal cost curve for supplier $i$.

    \item $p_i$: \pvcg{} payment to supplier $i$ from real curves.
    \item $\hat{p}_i$: \pvcg{} payment to supplier $i$ from interpolated curves.
\end{itemize}

\subsection{Game Setting}
In the CVGP game, a group of $n$ suppliers, denoted by $N = \{1,\ldots,n\}$, collaboratively produce a valuable virtual good consumed by  $m$ consumers. Agents can play dual roles, acting both as suppliers and consumers. On the supply side, these agents supply factors of production like data, labor, raw materials, computing power, and equipment. On the demand side, consumers consume the virtual good to derive utility.
This process is illustrated in Fig.~\ref{fig:teaser}.

Suppliers incur costs when providing factors of production. Let $x_i$ represent the procurement amount from supplier  $i$.  The cost for supplier $i$ can be described by $c(x_i, \gamma_i)$, where $\gamma_i$, referred to as the \emph{cost type}, characterizes the cost curve $c_i(\cdot)$. For now, consider $\gamma_i$ as representing the marginal costs $c_i'(x_i)$ over a sufficiently dense subset within the domain of $x_i$. We assume a capacity limit $\bar{x}_i$ for each supplier $i$.

By employing the Cr{\'e}mer-McLean mechanism to fully extract consumer surplus, the coordinator generates revenue denoted by $r(\bm{x})$. In this context, $r(\bm{x}) = \sum_{j=1}^m v(\bm{x}, \theta_j)$, where $\theta_j$ is termed as the \emph{valuation type}. The \emph{social surplus} $S$ is defined as the total valuation of consumers minus the total costs of suppliers. In our context:
\[
    S
    = \sum_{j=1}^m v(\bm{x}, \theta_j) - \sum_{i=1}^n c(x_i, \gamma_i)
    = r(\bm{x}) - \sum_{i=1}^n c(x_i, \gamma_i).
\]

At the outset of the game, suppliers report their capacity limits, denoted by $\hat{\bm{x}}$. During $t$th epoch of the \methodname{} iteration that will be detailed in Section~\ref{sec:rim}, supplier $i$ reports its marginal cost, $\hat{c'_i}^{(t)}$, at the procurement level $x_i^{(t)}$. The historical reported marginal costs of supplier $i$ are collectively referred to as $\hat{\gamma}_i$. Also, the coordinator measures the gradient of the revenue function, denoted by  $ \nabla r^{(t)}$, at the point $\bm{x}^{(t)}$.

\subsection{Assumptions}

Our theoretical analysis relies on the following assumptions about the revenue function and cost functions.

\begin{assumption}[Smoothness and Monotonicity]
    \label{asp:monotonicity}
    The revenue function $r(\bm{x})$ is smooth and monotonically increasing with respect to $\bm{x}$. The cost function $c(x_i, \gamma_i)$ is smooth and monotonically increasing with respect to $x_i$.
\end{assumption}

\begin{assumption}[No Impact at Zero Procurement]
    \label{asp:zeroProcurement}
     When the procurement level is zero, both revenue and costs are nullified:
    \[ r(\bm{0}) = 0, \quad c(0,\gamma_i) = 0, \quad \forall \gamma_i.\]
\end{assumption}

\begin{assumption}[Decreasing Cross Marginal Revenue]
    \label{asp:decreasingMarginalReturn}
    As other agents contribute more input resources, the marginal return from one agent's input factors decreases. Formally, for any \(i\in N\), and given \(x_i \ge x_i'\) and \(\bm{x}_{-i} \ge \bm{x}_{-i}'\), the following holds:
    \begin{align}
        &r(x_i,\bm{x}_{-i}) - r(x_i',\bm{x}_{-i}) 
        \le r(x_i,\bm{x}_{-i}') - r(x_i',\bm{x}_{-i}').
        \nonumber
    \end{align}
\end{assumption}

The justification for Assumption \ref{asp:decreasingMarginalReturn} is rooted in the law of diminishing marginal returns: as significant resources are already engaged in collaborative production, the additional marginal return from an extra unit of input factors tends to decline. It is worth noting that if the coordinator fully extracts the consumer surplus and the valuation functions satisfy:
\begin{align*}
    & v(x_i,\bm{x}_{-i}, \theta_j) - v(x_i',\bm{x}_{-i}, \theta_j) 
    \\
\le & v(x_i,\bm{x}_{-i}', \theta_j) - v(x_i',\bm{x}_{-i}', \theta_j), 
    \\
    & \qquad \forall j =1, ...,m, \, \theta_j,\, i \in N,\, x_i \ge x_i',\, \bm{x}_{-i} \ge \bm{x}_{-i}',
    \nonumber
\end{align*}
then Assumption~\ref{asp:decreasingMarginalReturn} holds true.
 
\begin{assumption}[Cr{\'e}mer-McLean Condition~\cite{kosenok2008individually, cremer-mclean-creemer1985optimal}]
    \label{asp:cremer}
    The prior belief about valuation types, Prior(\(\bm{\theta}\)), is identifiable and correlated.
\end{assumption}

This assumption guarantees full extraction of consumer surplus. Specifically: 
\begin{equation}\label{eqaution:WBBgivenCremer}
    r(\bm{x}) = \sum_{j=1}^{m}v_j(\bm{x}) := \sum_{j=1}^{m}v(\bm{x}, \theta_j).
    \nonumber
\end{equation}

\section{Theoretical Analysis}

\subsection{PVCG Sharing Rule}
The \pvcg{} payment to supplier $i$ is computed as:
\begin{equation}
    \label{eq:pvcgh}
    p_i(\hat{\bm{x}}, \hat{\bm{\gamma}}) 
    = S(\bm{x}^*,\hat{\bm{\gamma}})
    +c(x_{i}^{*},\hat{\gamma}_i) - h_i(\hat{\bm{x}}_{-i}, \hat{\bm{\gamma}}_{-i}),
\end{equation}
where $h_i(\cdot)$ is an arbitrary function and $\bm{x}^*$ maximizes the social surplus $S(\bm{x},\hat{\bm{\gamma}})$, based on reported cost types and reported capacity limits, i.e.:
\begin{align}
    % \label{equation:computeEta}
    \bm{x}^{*}(\hat{\bm{x}}, \hat{\bm{\gamma}}) & =
     \mathop{\textrm{argmax}}\limits_{
        \bm{x} \le \hat{\bm{x}}
        }
        S(\bm{x}, \hat{\bm{\gamma}})
    % \nonumber \\
    % &
    = \mathop{\textrm{argmax}}\limits_{
          \bm{x} \le \hat{\bm{x}}
        }
    \{ r(\bm{x})-\sum_{i=1}^{n}c(x_i , \hat{\gamma}_i)\}.
    \nonumber
\end{align}

Particularly, we can set:
\begin{equation}
    \label{eq:hform}
    h_i(\hat{\bm{x}}_{-i}, \hat{\bm{\gamma}}_{-i}) 
    := \mathcal{S}_{-i} (\hat{\bm{x}},\hat{\bm{\gamma}})
    = \mathcal{S}((0,\hat{\bm{x}}_{-i}),\hat{\bm{\gamma}})
    = S(\bm{z}_i^*,\hat{\bm{\gamma}}),
\end{equation}
where $\bm{z}_i^*$ maximizes the social surplus $S(\bm{x},\hat{\bm{\gamma}})$ based on reported cost types $\hat{\bm{\gamma}}$ and reported capacity limits $(0, \hat{\bm{x}}_{-i})$. 
Note that $z_{ii}^* \equiv 0$. Hence, $S(\bm{z}_i^*,\hat{\bm{\gamma}})$ does not depend on $\hat{\gamma}_i$ and $\hat{x}_i$. 

Consequently, the \pvcg{} payment becomes:
\begin{align} 
    \label{eq:pvcgPayment}
        p_i = & \mathcal{S}(\hat{\bm{x}},\hat{\bm{\gamma}})
    +c(x_{i}^{*},\hat{\gamma}_i)
    - \mathcal{S}((0,\hat{\bm{x}}_{-i}),\hat{\bm{\gamma}})
    \nonumber \\
    =  &
    \{ r(\bm{x}^*) - r(\bm{z}_{i}^*)\}
     -  
    \sum_{k=1, k\ne i}^{n}  \{ c(x_k^*, \hat{\gamma}_k) - c(z_{ik}^*, \hat{\gamma}_k)\}
   .
   % \nonumber
\end{align}

\subsection{Properties of PVCG} 
\label{sec:properties}
In this section, we will demonstrate that the \pvcg{} sharing rule in Eq.~\eqref{eq:pvcgPayment} simultaneously satisfies dominant-strategy incentive compatibility, Pareto efficiency, individual rationality, and weak budget balance. For definitions of these objectives, readers can refer to Appendix~\ref{appx:objectives} and standard game theory textbooks~\cite{mechanism_book_jackson2014mechanism,book_game_narahari2014game}. These objectives have also been incorporated into the IEEE Federated Machine Learning (P3652.1) Standard~\cite{ieee_standard_yang2021white}.

\subsubsection{Truthfulness}

First, we demonstrate that for an arbitrary function $h_i(\hat{\bm{x}}_{-i},\hat{\bm{\gamma}}_{-i})$, the \pvcg{} payment depicted in Eq.~\eqref{eq:pvcgh} incentivizes all suppliers to truthfully report their capacity limits and cost types.

\begin{proposition}[Dominant-strategy incentive compatibility on the supply side]
\label{proposition:IC}
    Given any supplier $i$, the dominant strategy on the supply side involves truthfully reporting its capacity limit \( \bar{x}_i \) and cost type \( \gamma_i \). Formally, this can be expressed as:
    \begin{align}
    \label{eq:DIC}
       & p_i(
            (\bar{x}_i,\hat{\bm{x}}_{-i}),
            (\gamma_i, \hat{\bm{\gamma}}_{-i})
            )
         - c(
                x^{*}_i(
                (\bar{x}_i,\hat{\bm{x}}_{-i}),
            (\gamma_i, \hat{\bm{\gamma}}_{-i})
            ),
                \gamma_i
        )
                \nonumber \\
        \ge &
        p_i(
            (\hat{x}_i,\hat{\bm{x}}_{-i}),
            (\hat{\gamma}_i, \hat{\bm{\gamma}}_{-i})
            )
         - c(
                x^{*}_i(
                (\hat{x}_i,\hat{\bm{x}}_{-i}),
            (\hat{\gamma}_i, \hat{\bm{\gamma}}_{-i})
            ),
                \gamma_i
        ),
        \nonumber \\
        & 
        \qquad \quad \qquad \qquad \qquad \qquad   \forall i\in N, \bar{x}_i, \gamma_i,
        \hat{x}_i,  \hat{\gamma}_i, \hat{\bm{x}}_{-i},
        \hat{\bm{\gamma}}_{-i}.
    \end{align}
\end{proposition}

Proof for Proposition~\ref{proposition:IC} is provided in Appendix~\ref{appx:proofIC}.

In our CVGP game, as illustrated in Fig.~\ref{fig:teaser}, an agent might act as both a supplier and a consumer. This dual role can cause the supply-side and demand-side problems to become intertwined. To tackle this complication, we assume full consumer surplus extraction, e.g., using the Cr{\'e}mer-McLean mechanism.

\begin{theorem}[Cr{\'e}mer-McLean Theorem~\cite{cremer-mclean-creemer1985optimal, mcafee1992correlated, kosenok2008individually}]
\label{theorem:CremerMcLean}
Given an arbitrary allocation rule for resources, there exists a mechanism that is Bayesian-Nash incentive-compatible, interim individually rational, and ex-post budget balanced. This mechanism can extract the full consumer surplus provided the Prior($\bm{\theta}$) is identifiable and the Cr{\'e}mer-McLean condition is satisfied.
\end{theorem}

An implementation algorithm for the Cr{\'e}mer-McLean mechanism can be found in \cite{cremer-computation-albert2015assessing}. Essentially, the Cr{\'e}mer-McLean Theorem indicates that the coordinator's revenue matches the full consumer surplus on the demand side:
\[
    r(\bm{x}) = \sum_{j=1}^m v(\bm{x}, \theta_j).
\]
Therefore, we deduce that an agent's behavior on the demand side will not influence their actions on the supply side, as their consumer surplus has been fully extracted, anyway.

\begin{corollary}
    \label{corollary:ICbothSide}
    By applying the \pvcg{} sharing rule on the supply side and implementing the Crémer-McLean mechanism on the demand side, truthfulness is ensured, even when an agent functions both as a consumer and a supplier.
\end{corollary}

\begin{proof}
    When agents act as both suppliers and consumers:

    (1) They have no incentive to deviate from truthfully reporting $(\bar{x}_i, \gamma_i)$ because any deviation brings no benefit on the demand side as the consumer surplus is fully extracted, but hurts the supply side due to the \pvcg{} sharing rule and Proposition ~\ref{proposition:IC}.

    (2) They also have no incentive to deviate from truthfully reporting $\theta_i$ because any deviation still leads to truth-telling regarding $(\bar{x}_i, \gamma_i)$ and hence, violates the Bayesian-Nash incentive compatibility due to the Cr{\'e}mer-McLean theorem.
\end{proof}

From Corollary \ref{corollary:ICbothSide}, it is established that the reported parameters should align with the true parameters, as expressed by:
\[
(\hat{\bm{x}}, \hat{\bm{\gamma}},\hat{\bm{\theta}}) = (\bar{\bm{x}}, \bm{\gamma},\bm{\theta}).
\]
This relationship plays a pivotal role in demonstrating the other properties of \pvcg{}.

\subsubsection{Pareto efficiency}

\begin{proposition}[Pareto efficiency] 
\label{proposition:pareto}
    When combined with the Cr{\'e}mer-McLean mechanism, \pvcg{} maximizes social surplus ex post.
\end{proposition}

Proof of ~\ref{proposition:pareto} is provided in Appendix~\ref{appx:proofPareto}.
To comprehend this, note that by implementing the \pvcg{} sharing rule, the utility of the suppliers are aligned with the social surplus.

\subsubsection{Individual rationality}

To ensure the properties of individual rationality (IR) and weak budget balance (WBB), the function $h_i(\hat{\bm{x}}_{-i}, \hat{\bm{\gamma}}_{-i})$ should be set to the maximum total surplus excluding supplier $i$, as depicted in Eq.~\eqref{eq:hform}.

\begin{proposition}[Ex-post individual rationality]
    \label{proposition:IR}
    As per the \pvcg{} sharing rule defined in Eq.~\eqref{eq:pvcgPayment}, all suppliers are guaranteed ex-post individual rationality. Specifically:
    \begin{align}\label{equation:IR}
        p_i (\hat{\bm{x}}, \hat{\bm{\gamma}}) 
         \ge
         c ( x_i^* (\hat{\bm{x}}, \hat{\bm{\gamma}})  , \gamma_i) , 
         \quad \forall i \in N.
    \end{align}
\end{proposition}

The proof for Proposition~\ref{proposition:IR} can be found in Appendix~\ref{appx:proofIR}.

\subsubsection{Weak budget balance}

The weak budget balance property further depends on Assumption~\ref{asp:decreasingMarginalReturn}, which assumes the concavity of the high-dimensional revenue function.

\begin{proposition}[Ex-post weak budget balance] 
    \label{proposition:WBB}
    If the assumption of decreasing cross marginal return is valid, then the \pvcg{} sharing rule given in Eq.~\eqref{eq:pvcgPayment} is ex-post weakly budget balanced.
\end{proposition}

Refer to Appendix~\ref{appx:proofWBB} for the proof of Proposition~\ref{proposition:WBB}.

\subsection{PVCG with Interpolated Curves}

In our prior theoretical discussions, we assumed the cost type  $\gamma_i$ provides sufficient information to precisely define the cost function  $c_i(\cdot)$. 
However, in practical scenarios, it is impractical to expect suppliers to report their entire cost curves. Instead, they might report their costs or marginal costs at selected observation points. Likewise, the coordinator can realistically only assess its revenue or revenue gradients at specific observation points.

Given this limitation, the coordinator interpolates the cost curves, $\hat{c}_i(\cdot)$, and the revenue curve, $\hat{r}(\cdot)$, based on these finite observations.

Using these interpolated curves, the coordinator can then compute the \pvcg{} payment as:
\begin{equation}
    \label{eq:paymentFitted}
    \hat{p}_i = \hat{r}(\hat{\bm{x}}^*)  - \hat{r} ( \hat{\bm{z}}_{i}^*) - \sum_{k=1 , k\ne i}^n [ \hat{c}_k (\hat{x}_i^*) - \hat{c}_k (\hat{z}_{ik}^*) ],
\end{equation}
where $\hat{\bm{x}}^*$ and $\hat{\bm{z}}_i^*$ maximize the interpolated social surplus function
$\hat{S}(\bm{x}) = \hat{r} (\bm{x}) - \sum_{i=1}^n \hat{c}_i (x_i)$. More specifically, we require:
\begin{align}
    \label{eq:optimalAllocation}
    \hat{\bm{x}}^{*} 
    = \mathop{\textrm{argmax}}\limits_{\bm{x} \le \hat{\bm{x}}} \hat{S}(\bm{x}),
    \quad
        \hat{\bm{z}}_i^{*} 
    = \mathop{\textrm{argmax}}\limits_{\bm{z} \le (0, \hat{\bm{x}}_{-i})} \hat{S}(\bm{z}),
    % \nonumber
\end{align}

A pivotal question emerges: 
\emph{Will \pvcg{} payments derived from interpolated curves retain the properties discussed previously?} 

The answer is affirmative, given that the interpolated curves and the real curves yield the coincident optimal allocation, i.e.:
\[
    \hat{\bm{x}}^{*} = \bm{x}^* .
\]

\begin{proposition}[PVCG with Interpolated Curves]
\label{proposition:fitcurve}
   Let us consider an interpolated revenue curve and a list of interpolated cost curves. If the following conditions hold:
   \begin{enumerate}
       \item Given truthfulness, the interpolated curves result in the same optimal allocation as the real curves.
       \item The interpolated curves always coincide with the real curves at the optimal allocation determined by the interpolated curves.
       \item Except at the interpolated optimal allocation, the interpolated cost curves lies below the real cost curves.
   \end{enumerate}
   Then, the \pvcg{} payments, determined by the interpolated curves, guarantee dominant-strategy incentive compatibility, thereby ensuring Pareto efficiency. Additionally, if the interpolated curves adhere to conditions specified in Assumptions~\ref{asp:monotonicity}-~\ref{asp:decreasingMarginalReturn},  the \pvcg{} payments also ensure individual rationality and weak budget balance.
\end{proposition}
Proof of this proposition is available in Appendix~\ref{appx:interpolation}.

\subsubsection{Circumvent the curse of dimensionality.}

When attempting to interpolate the revenue curve, we encounter the \emph{curse of dimensionality}. This phenomenon refers to the challenges posed by properly estimating a high-dimensional function with a limited set of observed values. To address this challenge, we employ the following specific form to interpolate the revenue function:
\[
    \hat{r}(\bm{x}) = \hat{\phi}( \bm{w}^T \bm{x}).
\]

In this representation, the high-dimensional function is simplified to the inner product of the input vector $\bm{x}$ and a weight vector $\bm{w}$, which is then passed through a univariate non-linear function $\hat{\phi}(\cdot)$. Given that $r(\cdot)$ is monotonically increasing, $\phi(\cdot)$ should also be increasing. With this structure in place, the coordinator's task becomes manageable: estimate the weight vector $\bm{w}$, and then interpolate the function $\hat{\phi}(\cdot)$ to obtain the desired revenue curve.

This dimension-reduction simplification can be implemented without compromising the desired properties of our \pvcg{} sharing rule, as supported by the findings in Proposition~\ref{proposition:fitcurve}.

\subsubsection{Learn the optimal allocation via gradient descent.}
\label{sec:gd}

To determine the optimal allocation using the interpolated curves, we employ the gradient descent method.

The gradient of $\hat{S}$ can be expressed as:
\[
    \nabla \hat{S} = \nabla \hat{r} - (\hat{c}_1', \cdots, \hat{c}_n') 
    =  \hat{\bm{w}} \hat{\phi}'(\hat{\bm{w}}^T\bm{x})  - \hat{\bm{c}'}(\bm{x}),
\]
where we use the bold notation $\hat{\bm{c}'}$ to represent the vector comprising the reported marginal costs from all suppliers.
Taking into account the reported capacity limits $\hat{\bm{x}}$, our update rule becomes:
\begin{equation}
\label{eq:updateRule}
\bm{x} \leftarrow \min\{ \max\{ \bm{x} + \alpha \nabla \hat{S} , \bm{0}\}, \hat{\bm{x}}\} ,
\end{equation}
where $\alpha$ represents the learning rate.

\subsubsection{Interpolate the derivative curves.}

To implement the gradient descent method, as shown in Eq.~\eqref{eq:updateRule}, the coordinator needs to know the derivatives of $\hat{\phi}(\cdot)$ and $\hat{c}_i(\cdot)$.
 Instead of interpolating these curves first and then determining their derivatives, a more accurate approach would involve interpolating the derivative curves---$\hat{\phi'}(\cdot)$ and $\hat{c_i'}(\cdot)$---directly. The integral of these derivative curves yields the original functions.

The \pvcg{} payment can be represented using definite integrals of the interpolated derivative curves:
\begin{align}
    \label{eq:pvcgFromDerivatives}
    \hat{p}_i 
    &
    = \hat{\phi} (\hat{\bm{w}}^T \hat{\bm{x}}^*)  - \hat{\phi} ( \hat{\bm{w}}^T \hat{\bm{z}}_{i}^*) - \sum_{k=1 , k\ne i}^n [ \hat{c}_k (\hat{x}_k^*) - \hat{c}_k (\hat{z}_{ik}^*) ]
    \nonumber \\ 
    & 
    = - \int_{\hat{\bm{w}}^T \hat{\bm{x}}^*}^{\hat{\bm{w}}^T \hat{\bm{z}}_{i}^*} \hat{\phi'}(s) ds
    + \sum_{k=1,k\ne i}^n \int_{\hat{x}_k^*}^{\hat{z}_{ik}^*} \hat{c'_k}(s)ds .
\end{align}

Thus, the coordinator can practically obtain the values of both marginal revenue and costs, interpolate these values to form the derivative curves, and subsequently use the integrals of these curves to calculate the \pvcg{} payments to suppliers.

\section{Methodology}
\label{sec:rim}

Let us revisit the gradient descent update rule as described in Eq.~\eqref{eq:updateRule}.  To locate the optimal  $\bm{x}^*$, the coordinator measures the gradient of its own revenue function  and collects the marginal costs across a sequence of allocations throughout the gradient descent trajectory:
\[
    \bm{x}^{(1)}, \bm{x}^{(2)}, ... \,, \bm{x}^{(t)}, ... 
\]
During this iterative interactive learning process, the coordinator gathers the reported marginal costs at these procurement levels:
\[
    \hat{\bm{c}'}^{(1)}, \hat{\bm{c}'}^{(2)}, ... \,, \hat{\bm{c}'}^{(t)}, ... 
\]
The coordinator also evaluates its marginal revenue at these points:
\[
    \nabla r^{(1)}, \nabla r^{(2)}, ... \,, \nabla r^{(t)}, ... 
\]
Leveraging this accumulated information, the coordinator can infer the gradient of the social surplus throughout its learning trajectory.
\[
    \nabla S^{(1)}, \nabla S^{(2)}, ... \,, \nabla S^{(t)}, ... 
\]
Such a methodology allows for determining the optimal procurement level via interactive learning with the suppliers.

\begin{figure}[h]
  \centering
  \includegraphics[width = \linewidth]{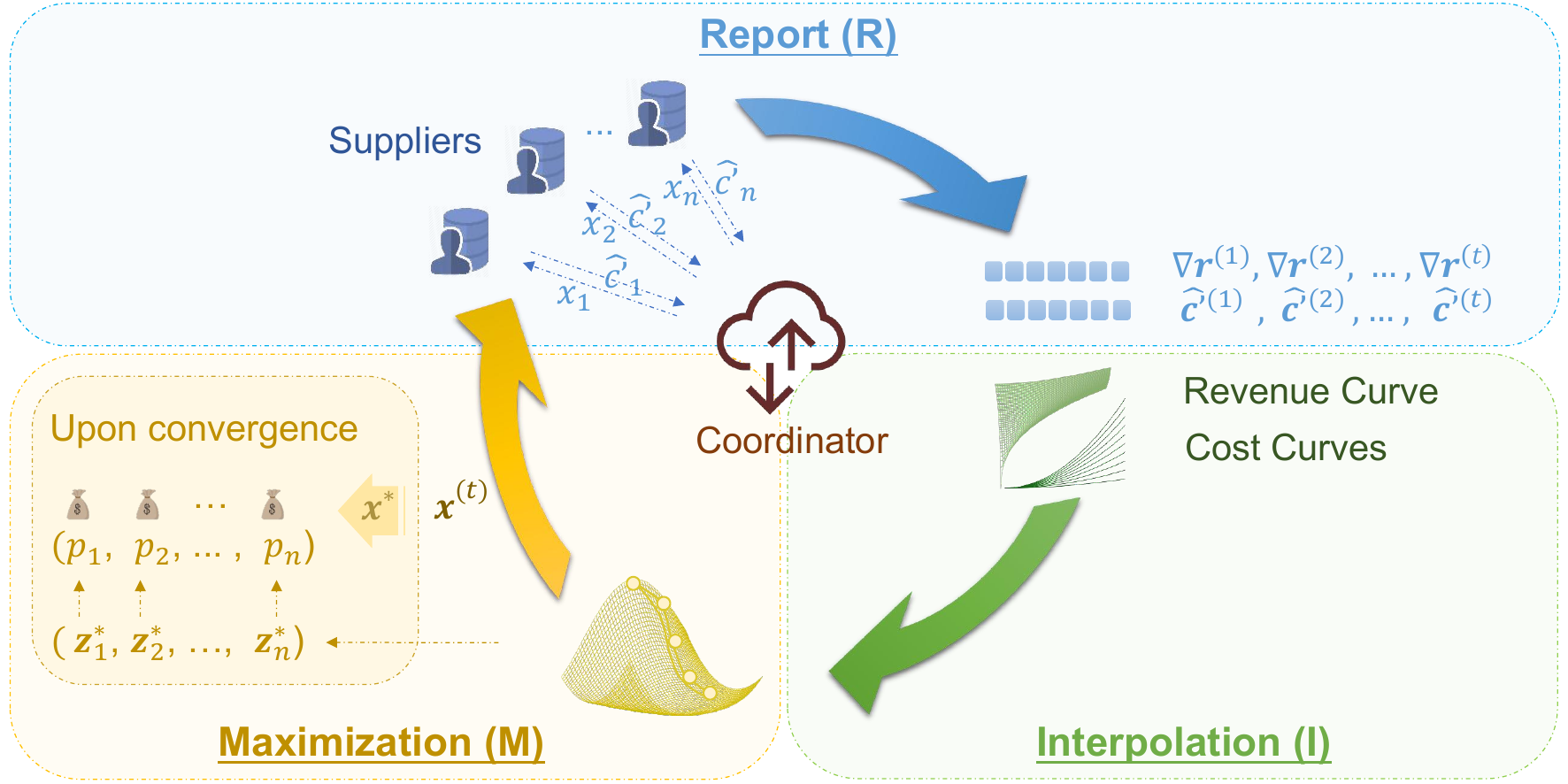}
  \caption{Report-Interpolation-Maximization iteration.}
  % \Description{The Report-Interpolation-Maximization Process.}
  \label{fig:rim}
\end{figure}

To compute the \pvcg{} payments as described in Eq.~\eqref{eq:pvcgPayment},  the coordinator must also identify  $n$ vectors $\bm{z}_1^*, ... , \bm{z}_n^*$ that maximize the total surplus in the absence of each supplier. Determining each $\bm{z}_i^*$  necessitates a unique learning trajectory for the coordinator. Expecting the suppliers to disclose all marginal costs across these multiple trajectories, resulting in a complexity of  $O(n^2)$, is impractical. To tackle this problem, the coordinator could interpolate both the marginal revenue curve and the marginal cost curves based on historical observations. In the following, we present the \underline{R}eport-\underline{I}nterpolation-\underline{M}aximization (\methodname{}) method in each epoch of interactive learning, as illustrated in Fig.~\ref{fig:rim}.

The \methodname{} method comprises three steps that are executed iteratively by the coordinator:
\begin{itemize}
    \item \textbf{R-step}: At the current procurement level, interact with suppliers to collect their marginal costs. Measure the revenue gradients at selected observation points.
    \item \textbf{I-step}: Based on historically observed revenue gradients and the reported marginal costs, interpolate both the marginal revenue and marginal cost curves.
    \item \textbf{M-step}: With the reported marginal costs and the observed revenue gradient, update the procurement level $\bm{x}^{(t)}$ by carrying out one update step of the gradient descent. 
    Utilizing the interpolated marginal revenue curve and marginal cost curves, the coordinator determines the optimal procurement level $\hat{\bm{x}}^{*(t)}$ that maximizes the interpolated social surplus. The iteration converges when the sequence of $\bm{x}^{(t)}$ stabilizes and becomes approximately equal to $\hat{\bm{x}}^{*(t)}$.
\end{itemize}
% The Protocol \pvcg{}, which employs the \methodname{} methodology, is elaborated upon in Algorithm~\ref{alg:pvcg}.

\subsection{Implication of Convergence}

The procurement level $\bm{x}$ is updated using gradients derived from real observations, ensuring that its stabilization mirrors the optimization of the true social surplus. Thus, when convergence is achieved, we have:
\[
    \bm{x}^{(t)} \simeq \bm{x}^*.
\]
The criteria for convergence also suggest that:
\[
    \bm{x}^{(t)} \simeq \hat{\bm{x}}^*.
\]
Here, $\hat{\bm{x}}^*$ represents the value that maximizes the social surplus based on the interpolated curves. Consequently, upon the \methodname{} iteration reaching convergence, it can be inferred:
\[
\bm{x}^* \simeq \hat{\bm{x}}^*.
\]
Additionally, our interpolation techniques, as explained below, ensure that all the premises of Proposition~\ref{proposition:fitcurve} are satisfied. Consequently, the \pvcg{} payment, as computed by Algorithm~\ref{alg:pvcg} in Appendix~\ref{appx:experiment}, maintains truthfulness along with other desired properties.

\subsection{Cost Curve Interpolation}

In interpolating cost and revenue curves, it is crucial to recognize that the real economy is noisy. Thus, we should select interpolation techniques that are robust against noise.

Given the inherent monotonic rise of the cost curve, it follows that its derivative is always non-negative. Hence, we model it as an exponential of a non-linear function, denoted as 
 $f_i$. Specifically:
\[
\log (1+ c_i'(x_i))  = f_i( \log(1+x_i))  .
\]
Such a logarithmic transformation is commonly employed in handling economic data. For instance, in econometrics, we often convert a stock price series into logarithmic returns.

During the $t$th epoch of the \methodname{} iteration, the coordinator has gathered $t$ observations on supplier $i$' marginal costs: 
\[
\hat{c_i'}^{(1)}, \hat{c_i'}^{(2)}, ... \,, \hat{c_i'}^{(t)} .
\]
From these observations, logarithmic transformations yield:
\[
\log(1+\hat{c_i'}^{(1)}), \log(1+ \hat{c_i'}^{(2)}), ... \,, \log(1+\hat{c_i'}^{(t)}).
\]
Employing these values, we derive an estimate for $\hat{f_i}(\cdot)$  via spline regression. \footnote{https://patsy.readthedocs.io/en/latest/spline-regression.html}

From this, the interpolated marginal cost is expressed as:
\[
\hat{c_i'}(x_i) = e^{\hat{f_i}(\log(1+x_i))} - 1.
\]

Lastly, to calculate the interpolated cost, we utilize the cost at a fixed point, 
$\hat{x}_i^*$, and integrate the interpolated marginal cost curve from this point:
\begin{equation}
\label{eq:costInterpolation}
\hat{c_i}(x_i) = c_i(\hat{x}_i^*) + \int_{\hat{x}_i^*}^{x_i} \hat{c_i'}(s) ds .    
\end{equation}

Observe that the interpolated cost curve, as given in  Eq.~\ref{eq:costInterpolation}, coincides with the real cost curve  $c_i(\cdot)$ at the point $\hat{x}_i^*$.
Due to the inherent nature of regression with noisy data, which often underestimates the slope, it is most likely that:
\[
    \Big| \frac{d}{dx_i} \hat{c_i'} (x_i) \Big| \le \Big| \frac{d}{dx_i} c_i'(x_i)\Big|, \quad 
    \forall x_i.
\]
Given that economic interpretation requires the real cost function to be convex,  this leads to:
\[
    \hat{c}_i(x_i) \le c_i(x_i), \, \quad \forall x_i \ne \hat{x}_i^*. 
\]
Consequently, the premises of Proposition~\ref{proposition:fitcurve} can be satisfied.

\subsection{Revenue Curve Interpolation}

As mentioned previously, when interpolating the revenue curve, it is essential to sidestep the so-called curse of dimensionality. We employ a specific functional form for the revenue function:
\[
    r (\bm{x}) = \phi( w_1 x_1 + w_2 x_2 + \cdots + w_n x_n ) = \phi(\bm{w}^T \bm{x}).
\]
Here, the weight vector  $\bm{w}$ and the univariate function $\phi(\cdot)$ both need to be estimated from observed data.
Given the revenue function's monotonicity, $\phi(\cdot)$ should also be increasing, and hence the weight vector $\bm{w}$ should be non-negative.

Considering the revenue gradient:
\begin{align}
    \label{eq:revGrad}
    \nabla r (\bm{x}) = \phi'(\bm{x}) \bm{w} \ge \bm{0}.
\end{align}
By the $t$th epoch of the \methodname{} iteration, the coordinator has collected observations of the revenue gradients at points:
\[
    \bm{x}^{(1)}, \bm{x}^{(2)}, ...\,, \bm{x}^{(t)}. 
\]
Let us define:
\begin{align*}
&
\delta_{i\tau} := \frac{\partial r}{\partial x_i} \Big|_{\bm{x}^{(\tau)}}, \quad 
\psi_{\tau} :=\phi' (\bm{w}^T\bm{x}^{(\tau)}), 
\\
& \qquad \quad i = 1, ...\,,n, \, \tau = 1, ...\,, t.     
\end{align*}

Then, the relationship in Eq.~\eqref{eq:revGrad} becomes:
\[
    \delta_{i\tau} = w_i \psi_{\tau},\quad i = 1,..., n, \, \tau = 1,..., t.
\]
In matrix form, this becomes the outer product of two vectors:
\[
    \bm{\delta} \simeq \bm{w} \otimes \bm{\psi}.
\]

Given that economic interpretations require $\bm{w} \ge \bm{0}$ and $\bm{\psi} \ge \bm{0}$, we employ non-negative matrix factorization (NMF)
to estimate $\hat{\bm{w}}$ and $\hat{\bm{\psi}}$ based on the observed matrix $\bm{\delta}$.
\footnote{https://scikit-learn.org/stable/modules/generated/sklearn.decomposition.
NMF.html} 

With the estimated weight vector $\hat{\bm{w}}$, we are able to compute: 
\[
    y^{(\tau)} := \hat{\bm{w}}^T \bm{x}^{(\tau)}, \quad \tau = 1, 2, ..., t.
\]
Based on the definition of $\hat{\psi}_{\tau}$, we know:
\[
     \phi'(y^{(\tau)}) \simeq \hat{\psi}_{\tau}, \quad \tau = 1, 2, ..., t.
\]
Hence, we have obtained $t$ observations on the derivative of the non-negative function $\phi'(\cdot)$.
This enables us to interpolate the curve $\hat{\phi'}(\cdot)$ in a manner similar to the method detailed in the previous subsection.

Once $\hat{\phi'}(\cdot)$ is obtained, revenue at any point $\bm{x}$ can be determined using the revenue at a fixed point $\hat{\bm{x}}^*$ and the integral of $\hat{\phi'}(\cdot)$:
\begin{align*}
    \hat{r}(\bm{x}) 
    & = r(\hat{\bm{x}}^*) + [ \hat{\phi}(\hat{w} \bm{x}) - 
    \hat{\phi}(\hat{w} \hat{\bm{x}}^*) ] 
     \\
     & = r(\hat{\bm{x}}^*) + \int_{\hat{w} \hat{\bm{x}}^*}^{\hat{w} \bm{x}} \hat{\phi}'(s)ds.
    \nonumber
\end{align*}
It is worth noting that the interpolated revenue curve coincides with the real curve at the point $\hat{\bm{x}}^*$.
    
\subsection{Optimization on Interpolated Curves}

To determine $\hat{\bm{z}}_i^*$ as per Eq.~\eqref{eq:optimalAllocation}, gradient descent is applied to the interpolated social surplus function $\hat{S}(\cdot)$. The gradient at any given point $\bm{z}$ is:
\[
    \nabla \hat{S} (\bm{z}) 
    = \hat{\phi'}(\hat{\bm{w}}^T \bm{z})\hat{\bm{w}} -
    (\hat{c_1'}(z_1), ..., \hat{c_n'}(z_n) ).
\]
Utilizing the interpolated curves, optimal allocations can be located via the gradient descent approach detailed in Subsection~\ref{sec:gd}

After determining $\bm{z}_i^*$ for each $i\in N$, \pvcg{} payments are calculated using the definite integrals of the interpolated derivative curves, as outlined in Eq.~\eqref{eq:pvcgFromDerivatives}.

\section{Experiment}

An exhaustive experimental study of \pvcg{} is beyond the scope of this paper.
Our exploratory experiment is limited to addressing the following research questions:

\begin{itemize}
    \item \textbf{RQ1.} Can the \methodname{} method determine the optimal procurement level in a noisy economic environment?
    \item \textbf{RQ2.} Is the \methodname{} method capable of accurately computing the \pvcg{} payments?
    \item \textbf{RQ3.} How do the payments to suppliers fluctuate based on their capacity limits and cost coefficients?
    \item \textbf{RQ4.} Is the \pvcg{} sharing rule \emph{fairness aware}, ensuring a comparable unit price to all suppliers?
\end{itemize}

\subsection{Experimental Design}

% While economics typically leans more toward the empirical rather than the experimental,

We carry out the experiment within a simulated toy economy to evaluate the efficacy of our \methodname{} method.  We envision the economy as a black-box system driven by certain underlying functions. Agents in the economy may pose noisy queries to economic oracles.

\begin{figure}[h]
  \centering
  \includegraphics[width = \linewidth]{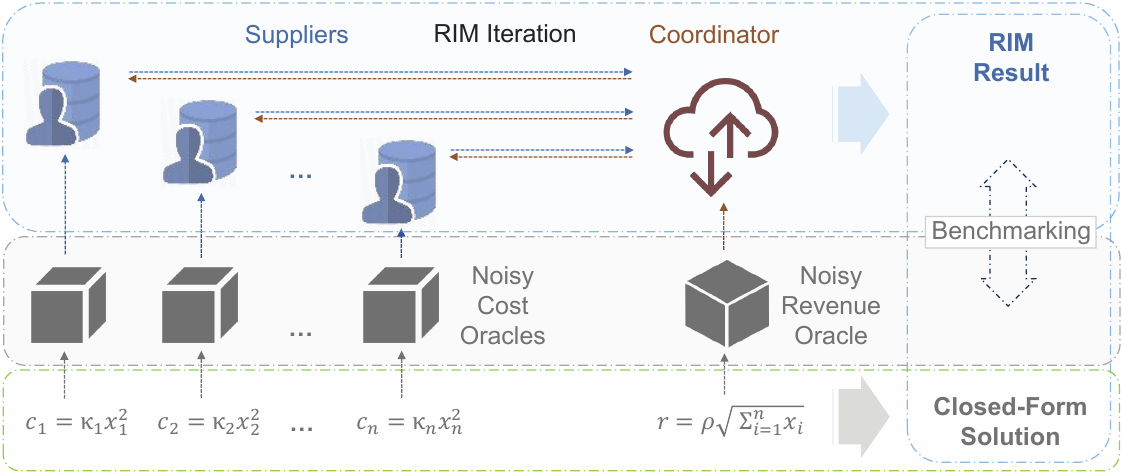}
    % \vspace{-0.5cm}
  \caption{Experimental design.}
  % \Description{The Experimental Design for the Report-Interpolation-Maximization Method.}
    % \vspace{-0.2cm}
\end{figure}

We posit that the economic oracles operate under the subsequent revenue and cost functions:
\begin{equation}
% \label{eq:realCurve}
    r(\bm{x}) = \rho \sqrt{\sum_{i=1}^n x_i} \,, \quad c_i(x_i) = \kappa_i x_i^{2}, \,\, i\in N.
    \nonumber
\end{equation}
While agents remain unaware of these functions' precise formulations, they have noisy oracle access to the revenue gradients and marginal costs,  echoing the observation process in a real economy.  They then interpolate the curves based on the query outcomes.

In our simulated economy, the socially optimal allocation can be calculated through the following closed-form solution:

\[
    x_i^* = \max\{\frac{C}{\kappa_i},  \bar{x}_i \}, \quad \textrm{where} \,\,  C= \frac{\rho}{ 4  \sqrt{\sum_{i=1}^n x_i^{*}}}.
\]
Values derived from this closed-form solution (which are unknown to the agents) serve as benchmarks for evaluating our \methodname{} method.

We consider a scenario with  $10$ suppliers, each having a capacity limit that ranges from $1.0$ to $10.0$ and a corresponding cost coefficient $\kappa_i$ with values from $1.0$ to $10.0$. The marginal costs and revenue gradients are observed with a noise level of  $10\%$. 
% We employ a learning rate of $0.1$ and a momentum coefficient of $0.9$ in interactive gradient descent epoch. 
Further experimental details are provided in Appendix~\ref{appx:experiment}. 

\subsection{Results and Interpretation}

\begin{figure}[h]
  \centering
  \includegraphics[width= \linewidth]{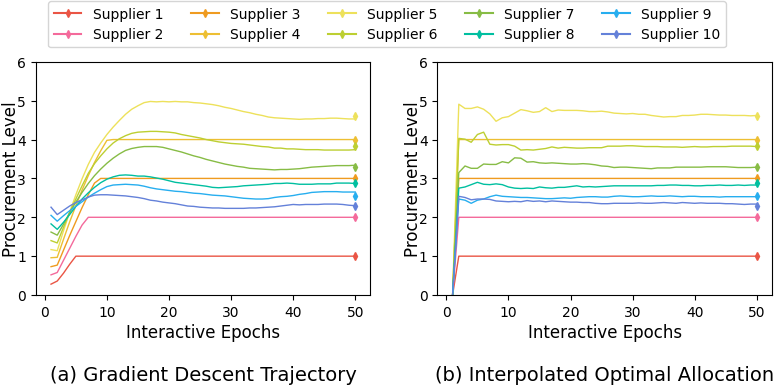}
    % \vspace{-0.5cm}
  \caption{Trajectory of the RIM iteration. Diamond markers highlight the values obtained from the closed-form solution (addressing RQ1).}
  % \Description{Trajectory of RIM iterations. Diamond markers highlight the values obtained from the closed-form solution, Addressing RQ1.}
  \label{fig:learning}
\end{figure}

The \methodname{} iteration observed convergence of the procurement levels after $50$ interactive epochs.
The trajectory is illustrated in Figure ~\ref{fig:learning}. Notably, the diamond markers signify the optimal procurement levels derived from the closed-form solution.

\begin{figure}[h]
  \centering
  \includegraphics[width= \linewidth]{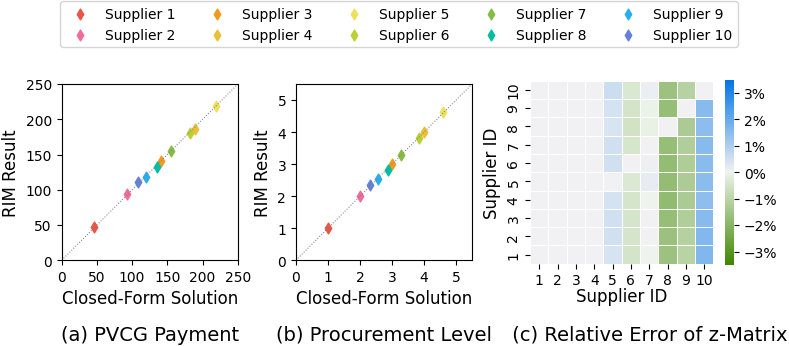}
    % \vspace{-0.5cm}
  \caption{Comparison of RIM results with closed-form solution values (addressing RQ2).}
  % \Description{Comparison of RIM Results with Values Obtained from Closed-Form Solution Values, Addressing RQ2.}
  \label{fig:scatter}
\end{figure}

In Figure \ref{fig:scatter}, we compare the \methodname{} results for \pvcg{} payments and optimal procurement levels against values from the closed-form solution, highlighting \methodname{}'s accuracy.
% However, in our experiments, it slightly underestimates the payment for supplier $6$ and overestimates for supplier $7$. These discrepancies can be attributed interpolation errors related to the observation noise. 

\begin{figure}[h]
  \centering
  \includegraphics[width= \linewidth]{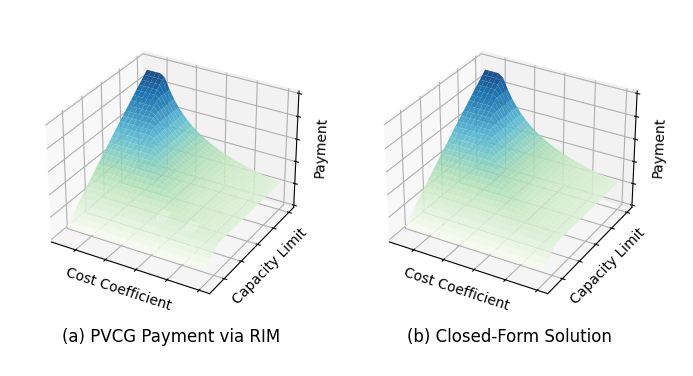}
  % \vspace{-0.7cm}
  \caption{Variation in payment to supplier $5$: Influence of the capacity limit and the cost coefficient (addressing RQ3).}
  % \Description{Variation of Payment to Supplier $5$ Based on Capacity Limit and Cost Parameter Influences, Addressing RQ3.}
  \label{fig:surface}
  % \vspace{-0.3cm}
\end{figure}

The overall trend of \pvcg{} payments using \methodname{} closely aligns with those derived from the closed-form solution. As an example, we show the payment to supplier $5$, as illustrated in Figure~\ref{fig:surface}:
\begin{enumerate}
    \item When a supplier's cost coefficient is low, the procurement level typically reaches its capacity limit, resulting in a payment that is directly proportional to the capacity limit.
    \item  On the other hand, with higher cost coefficients, the procurement level lies at the point where the supplier's marginal cost matches their contribution's marginal revenue. The payment is influenced correspondingly.
\end{enumerate}
% This fact answers why the procurement level in Subfigure (b) of Figure~\ref{fig:scatter} still aligns well with the theoretical values.

% \vspace{-0.5cm}

\begin{figure}[h]
  \centering
  \includegraphics[width= \linewidth]{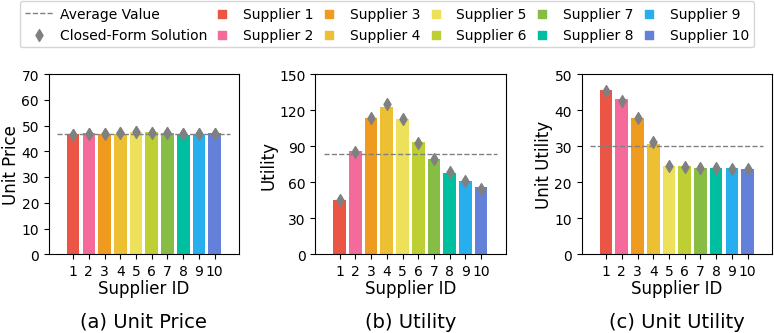}
  \caption{Unit price, utility, and unit utility across suppliers (addressing RQ4).}
  % \Description{Depiction of Unit price, Utility, and Unit Utility across Suppliers, Addressing RQ4.}
  \label{fig:unit}
  % \vspace{-0.5cm}
\end{figure}

\paragraph{Fairness aware.} The unit prices awarded to suppliers are delineated in Figure~\ref{fig:unit}. In this figure, the dashed line represents the average unit price across suppliers, while the diamond markers denote values obtained from the closed-form solution. Notably, despite the variance in payment magnitudes, the unit prices manifest remarkable consistency. An additional observation is that as the cost coefficients for suppliers differ, the unit utility---defined as the difference between the unit price and the unit cost---exhibits a decline in correlation with the increased cost coefficients.

Our result indicates that the \pvcg{} sharing rule embodies the ethos of \emph{equal pay for equal work}. To understand this,  recall that the \pvcg{} payment to supplier $i$ is determined based on parameters reported by all suppliers excluding supplier $i$. As the number of suppliers increases, the impact of each individual supplier gravitates toward a mean value.

\section{Conclusion}

We introduce the \pvcg{} sharing rule, envisioning Pareto-efficient collaborative production steered by a benevolent AI central planner.  
Mathematical proofs confirm that \pvcg{} ensures truthfulness, Pareto efficiency, individual rationality, and weak budget balance. 
% These properties are consistent with the \pvcg{} payments derived from interpolated curves. 
% To address the computational complexities arising from a continuous strategy space, we introduce the \methodname{} method. This approach effectively computes the \pvcg{} sharing rule, even amidst economic noise.
Our \methodname{} method effectively computes the \pvcg{} sharing rule, even admist economic noise.

% \clearpage



%\bibliographystyle{IEEEtran}
\bibliographystyle{plain}
\bibliography{mybib.bib}

\appendices
\section{Background}

\subsection{Pareto Efficiency}

In an economy,  a \emph{Pareto improvement} is a change that makes some individuals better off without disadvantaging others. An allocation that precludes further Pareto improvements is called \emph{Pareto efficient}.
Pareto efficiency aligns with the maximization of the \emph{social surplus}, which contains consumer surplus and producer surplus. 
% Social surplus represents the total value consumers attribute to goods, subtracting the total production costs borne by producers.

\begin{theorem}[The First Theorem of Welfare Economics]
    Given complete markets with full information and perfect competition, the economic equilibrium is Pareto efficient.
\end{theorem}

This theorem is based on two assumptions:
\begin{enumerate}
    \item In the economy, all goods are \emph{excludable} and \emph{rivalrous}. 
    % Here, a good is  excludable if  its consumption is restricted to paying customers. A good is rivalrous  if its consumption by one consumer prevents simultaneous consumption by others. 
    \item 
    % For all goods, there exists a \emph{perfectly competitive market}. In this market model,
    Participants are price takers, obligated to accept the prevailing market prices.
\end{enumerate}

Neither of these assumptions holds true for virtual goods.

\subsection{Free Rider Problem}

In economics, we categorize the virtual good as a \emph{club good}. A club good is excludable but non-rivalrous. Here, a good is non-rivalrous when the cost to serve an additional user is zero.

To understand the concept of club good, imagine a club pooling funds to purchase a TV for communal use. Each member might hope someone else bears the cost, knowing they would benefit fully once the TV is acquired. Consequently, everyone might attempt to pay as little as possible. Economists describe this phenomenon as individuals trying to \emph{free ride} on one another.

Economists have proposed the VCG mechanism to address the free rider problem.

\subsection{Objectives of Mechanism Design}
\label{appx:objectives}

\emph{Mechanism design}, rooted in game theory, aims to achieve a set of objectives. Below are some of the standard goals:

\paragraph{Incentive compatibility (IC)}
 % Incentive-compatibility implies truthfulness. 
 This includes dominant-strategy incentive compatibility (DSIC), where truth-telling is always optimal, and Bayesian-Nash incentive compatibility (BNIC), where truth-telling is optimal when others also act truthfully.

\paragraph{Allocative efficiency} 
Resources are optimally allocated among participants to attain Pareto efficiency.

\paragraph{Individual rationality (IR)}

Participants' payoff can cover their costs.

\paragraph{Budget balance (BB)}
Strong budget balance (SBB) means that the coordinator's incoming and outgoing payments are equal. Weak budget balance (WBB) allows for potential profits but not losses for the coordinator.

\subsection{VCG Recap}
\label{sec:vcgRecap}

\begin{figure}[h]
  \centering
  \includegraphics[width=\linewidth]{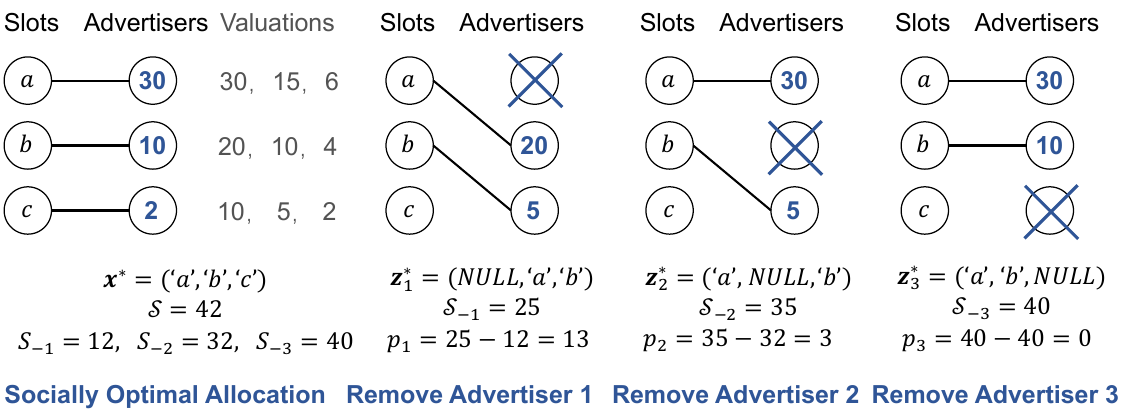}
  \caption{Example of the VCG Ad Auction.}
  % \Description{The Report-Interpolation-Maximization Process.}
  \label{fig:vcg}
\end{figure}

To elucidate the VCG mechanism, we borrow an example from~\cite{facebook-profiting-leo2020vcg}. This example showcases the approach taken by Facebook ad auctions. In Figure ~\ref{fig:vcg}, advertisers compete for advertising slots (`a', `b', `c'), each having a distinct click-through rate.

In this setup, we denote the \emph{allocation} of slots by the vector $\bm{x}$. The $i$th element of $\bm{x}$ indicates the slot id allocated to the $i$th advertiser. The set of all feasible allocations is represented by $X$. $Z_i$ denotes the set of all feasible allocations excluding advertiser $i$.

The VCG mechanism is a generalization of the \emph{second-price sealed-bid auction}. 
In essence, the VCG payment is the surplus gain of other advertisers resulting from the removal of advertiser $i$:
\begin{equation}
    % \label{eq:second-price}
    - t_i = S_{-i}(\bm{x}^*) - S_{-i}(\bm{z}_{i}^*)  .
\nonumber
\end{equation}
In this context, the negative sign before $t_i$ signifies charges rather than payments to the advertisers. 
The symbol $\bm{x}^*$  denotes the optimal slot allocation with all advertisers included, 
while $\bm{z}_i^*$ signifies the optimal slot allocation excluding the $i$th advertiser, i.e.:
\[
    \bm{x}^* = \mathop{\textrm{argmax}}\limits_{\bm{x} \in X} S(\bm{x}), \quad
    \bm{z}_i^* =\mathop{\textrm{argmax}}\limits_{\bm{z} \in Z_i} S_{-i}(\bm{z}).
\]
This VCG payment formula can be tailored to fit our CVGP framework that functions in a reverse auction setting with capacity limits and a continuous strategy space.
% Consequently, we derive the \pvcg{} sharing rule.

\section{Proofs}

\subsection{Proof of Proposition~\ref{proposition:IC}}
\label{appx:proofIC}

\begin{proof}[Dominant-Strategy Incentive Compatibility]
Based on the reported values of $\hat{\bm{x}}$ and $\hat{\bm{\gamma}}$, the coordinator determines the socially optimal allocation $\bm{x}^*(\hat{\bm{x}}, \hat{\bm{\gamma}})$. We distinguish between scenarios: one 
where $x^*_i(\hat{\bm{x}},\hat{\bm{\gamma}}) $ remains within supplier $i$'s capacity limit, and another where it exceeds the limit.

Case 1: If $x^*_i(\hat{\bm{x}},\hat{\bm{\gamma}}) > \bar{x}_i$, supplier $i$ cannot provide the allocated input resource $x^*_i(\hat{\bm{x}},\hat{\bm{\gamma}})$ as it surpasses its capacity limit. Under this circumstance,  supplier $i$ faces a penalty, potentially losing the payments for prior contributions.
Hence, the RHS of  Eq. \eqref{eq:DIC} becomes negative, ensuring that Eq. \eqref{eq:DIC} is satisfied. Therefore, a rational supplier would avoid reporting $\hat{x}_i$ that results in $x^*_i(\hat{\bm{x}},\hat{\bm{\gamma}}) > \bar{x}_i$.

Case 2: If $x^*_i (\hat{\bm{x}},\hat{\bm{\gamma}}) \le \bar{x}_i$, we will demonstrate that, by truthfully reporting $\bar{x}_i$ and $\gamma_i$, supplier $i$'s utility is not diminished.

By definition, we have:
\[
    \bm{x}^* (\hat{\bm{x}},\hat{\bm{\gamma}}) \le (\hat{x}_i, \hat{\bm{x}}_{-i}).
\]
This inequality and the condition $x^*_i (\hat{\bm{x}},\hat{\bm{\gamma}}) \le \bar{x}_i$ ensure that:
\[
    \bm{x}^*(\hat{\bm{x}},\hat{\bm{\gamma}}) \le (\bar{x}_i , \hat{\bm{x}}_{-i}).
\]

Since $\mathcal{S}(( \bar{x}_i, \hat{\bm{x}}_{-i}), (\gamma_i, \hat{\bm{\gamma}}_{-i}))$ is defined as the maximum social surplus given $( \bar{x}_i, \hat{\bm{x}}_{-i})$ and $(\gamma_i, \hat{\bm{\gamma}}_{-i})$, we can deduce:
\begin{align}
    \label{eq:proofDIC1}
    & \mathcal{S}( (\bar{x}_i, \hat{\bm{x}}_{-i}), (\gamma_i, \hat{\bm{\gamma}}_{-i}))
    \nonumber \\
    \ge & 
    r(\bm{x}^*(\hat{\bm{x}},\hat{\bm{\gamma}}))
   - \sum_{k=1, k\ne i}^{n}c(x_k^*(\hat{\bm{x}},\hat{\bm{\gamma}}), \hat{\gamma}_k) - c(x_i^*(\hat{\bm{x}},\hat{\bm{\gamma}}), \gamma_i)
    \nonumber \\
    = &
    S(\bm{x}^*(\hat{\bm{x}},\hat{\bm{\gamma}}), \hat{\bm{\gamma}}) + c(x_i^*(\hat{\bm{x}},\hat{\bm{\gamma}}), \hat{\gamma}_i) - c(x_i^*(\hat{\bm{x}},\hat{\bm{\gamma}}), \gamma_i)
    \nonumber \\
    = &
    \mathcal{S}(\hat{\bm{x}},\hat{\bm{\gamma}}) + c(x_i^*(\hat{\bm{x}},\hat{\bm{\gamma}}), \hat{\gamma}_i) - c(x_i^*(\hat{\bm{x}},\hat{\bm{\gamma}}), \gamma_i)
    .
    % \nonumber
\end{align} 

Thus, we can express:
\begin{align}
\label{eq:proofDIC2}
& \mathcal{S}( (\bar{x}_i, \hat{\bm{x}}_{-i}), (\gamma_i, \hat{\bm{\gamma}}_{-i}))
+ c(x_i^* ((\bar{x}_i, \hat{\bm{x}}_{-i}), (\gamma_i, \hat{\bm{\gamma}}_{-i})), \gamma_i)
\nonumber\\
& \quad \quad
- c(x_i^* ((\bar{x}_i, \hat{\bm{x}}_{-i}), (\gamma_i, \hat{\bm{\gamma}}_{-i})), \gamma_i)
\nonumber
\\
\ge &
\mathcal{S}(\hat{\bm{x}},\hat{\bm{\gamma}}) + c(x_i^*(\hat{\bm{x}},\hat{\bm{\gamma}}), \hat{\gamma}_i)
- c(x_i^*(\hat{\bm{x}},\hat{\bm{\gamma}}), \gamma_i).
\end{align}

By adding $h_i(\hat{\bm{x}}_{-i},\hat{\bm{\gamma}}_{-i})$
to both sides of Eq. \eqref{eq:proofDIC2} and upon substituting with:
\[p_i(\hat{\bm{x}},\hat{\bm{\gamma}})  =\mathcal{S}(\hat{\bm{x}},\hat{\bm{\gamma}}) + c ( x_i^*(\hat{\bm{x}},\hat{\bm{\gamma}}), \hat{\gamma}_i) + h_i(\hat{\bm{x}}_{-i},\hat{\bm{\gamma}}_{-i}),
\] 
we arrive at Eq. \eqref{eq:DIC}.
\end{proof}

\subsection{Proof of Proposition~\ref{proposition:pareto}}
\label{appx:proofPareto}

\begin{proof}[Pareto Efficiency.]
    Given the incentive compatibility, suppliers are ensured to truthfully report their private parameters such that  $\hat{\bm{x}} = \bar{\bm{x}}$ and $\hat{\bm{\gamma}}=\bm{\gamma}$.
    Consequently, the allocation achieved from \pvcg{} becomes:
    \[
        \bm{x}^*(\hat{\bm{x}}, \hat{\bm{\gamma}}) = \bm{x}^*(\bar{\bm{x}}, \bm{\gamma}).
    \]

    From the definition of $\bm{x}^*(\bar{\bm{x}}, \bm{\gamma})$, we know that:
    \begin{align} 
    \label{eq:proofPareto1}
    & r(\bm{x}^*(\bar{\bm{x}}, \bm{\gamma}))
    - \sum_{i=1}^n c(x_i^*(\bar{\bm{x}}, \bm{\gamma}), \gamma_i)
    \nonumber \\
    \ge & r(\bm{x}) - \sum_{i=1}^n c(x_i, \gamma_i),\quad \forall \bm{x} \le \bar{\bm{x}}
    .
    % \nonumber
    \end{align}

    Given that  the Cr{\'e}mer-McLean mechanism extracts the full consumer surplus on the demand side, we can infer:
    \[
        r(\bm{x}^{*}(\bar{\bm{x}}, \bm{\gamma})) = \sum_{j=1}^m v(\bm{x}^{*}(\bar{\bm{x}}, \bm{\gamma}), \theta_j).
    \]
    Substituting this into Eq.~\eqref{eq:proofPareto1}, we have:
    \begin{align} 
    % \label{eq:proofPareto2}
    & S(\bm{x}^{*}(\bar{\bm{x}}, \bm{\gamma}), \bm{\gamma}) = \sum_{j=1}^m v(\bm{x}^{*}(\bar{\bm{x}}, \bm{\gamma}), \theta_j)
    - \sum_{i=1}^n c(x_i^*(\bar{\bm{x}}, \bm{\gamma}), \gamma_i)
    \nonumber \\
    & \ge \sum_{j=1}^m v(\bm{x}, \theta_j) - \sum_{i=1}^n c(x_i, \gamma_i)
    = S(\bm{x}, \bm{\gamma}), \quad \forall \bm{x} \le \bar{\bm{x}}
    .
    \nonumber
    \end{align}
    This inequality implies that the \pvcg{} allocation, $\bm{x}^{*}(\bar{\bm{x}}, \bm{\gamma})$, achieves the maximal ex-post social surplus under the true parameters.
\end{proof}

\subsection{Proof of Proposition~\ref{proposition:IR}}
\label{appx:proofIR}

\begin{proof}[Individual Rationality]
    Based on the established incentive compatibility, we have:
    \[h_i(\hat{\bm{x}}_{-i}, \hat{\bm{\gamma}}_{-i})
    :=   \mathcal{S}((0,\hat{\bm{x}}_{-i}),\hat{\bm{\gamma}})
    =   \mathcal{S}((0,\bar{\bm{x}}_{-i}),\bm{\gamma}). 
    \]
    Then, the net utility for supplier $i$ (i.e., payment minus cost) is expressed as:
\begin{align}
    % \label{equation:IRproof}
    &p_i(\hat{\bm{x}},\hat{\bm{\gamma}})
    - c(x_{i}^{*}(\hat{\bm{x}},\hat{\bm{\gamma}}),\gamma_i)
    \nonumber \\
    = &
    p_i(\bar{\bm{x}},\bm{\gamma})
    - c(x_{i}^{*}(\bar{\bm{x}},\bm{\gamma}),\gamma_i)
    \nonumber \\
    = &
    \mathcal{S}(\bar{\bm{x}},\bm{\gamma}) + c(x_{i}^{*}(\bar{\bm{x}},\bm{\gamma}),\gamma_i)
        - \mathcal{S}((0,\bar{\bm{x}}_{-i}),\bm{\gamma}) - c(x_{i}^{*}(\bar{\bm{x}},\bm{\gamma}),\gamma_i)
    \nonumber \\
    = &
    \mathcal{S}(\bar{\bm{x}},\bm{\gamma}) 
        - \mathcal{S}((0,\bar{\bm{x}}_{-i}),\bm{\gamma})  \ge 0 \quad \qquad
        \textrm{by definition of }    \mathcal{S}(\bar{\bm{x}},\bm{\gamma}). 
    \nonumber
\end{align}
Thus, the ex-post payment to each supplier $i$ always covers its cost.
\end{proof}

\subsection{Proof of Proposition~\ref{proposition:WBB}}
\label{appx:proofWBB}

\begin{proof}[Weak Budget Balance]
    By leveraging the established truthfulness,
    we derive the ex-post total payment to all suppliers as:
    \begin{align}
    \label{eq:proofWBB0}
        &\sum_{i=1}^{n}p_i(\bar{\bm{x}},\bm{\gamma}) 
        % \nonumber \\
        % &
        =\sum_{i=1}^{n}[\mathcal{S}(\bar{\bm{x}},\bm{\gamma}) 
        + c(x_{i}^{*}(\bar{\bm{x}},\bm{\gamma}),\gamma_i)
        - \mathcal{S}((0,\bar{\bm{x}}_{-i}),\bm{\gamma})].
        % \nonumber
    \end{align}
   
Considering the definition of
$\mathcal{S}((0, \bar{\bm{x}}_{-i}),  \bm{\gamma}, \bm{\theta} )$, we deduce the following inequality:
\begin{align}
    % \label{equation:socialSurplusCompare2}
    & \mathcal{S}((0, \bar{\bm{x}}_{-i}),\bm{\gamma} )
    \ge
    S(\bm{z} ,\bm{\gamma})
    ,\quad \forall \bm{z} \le (0, \bar{\bm{x}}_{-i}).
    \nonumber
\end{align}
This inequality is valid for $\bm{z} = (0, \bm{x}^*_{-i}(\bar{\bm{x}}, \bm{\gamma}))$. Specifically:
\begin{align}
\label{eq:proofWBB1}
& \mathcal{S}((0, \bar{\bm{x}}_{-i}),\bm{\gamma} )
\ge 
S((0,\bm{x}^*_{-i}(\bar{\bm{x}}, \bm{\gamma})),\bm{\gamma}).
% \nonumber
\end{align}

Drawing from Assumptions ~\ref{asp:zeroProcurement} and ~\ref{asp:decreasingMarginalReturn}, 
we also have:
\begin{align}
\label{eq:proofWBB2}
& r(\bm{x}^{*})
=
r(\bm{x}^{*}) -  r(\bm{0})  
\qquad \qquad \qquad \quad  \textrm{by Assumption~\ref{asp:zeroProcurement}} 
\nonumber \\
=
& \sum_{i=1}^n \{ r(x_1^*, \cdots, x_{i-1}^*  , x_i^* , \bm{0}) 
 -  
r(x_1^*, \cdots, x_{i-1}^*  , 0 , \bm{0}) \}
\nonumber \\
\ge
& \sum_{i=1}^n \{ r(x_1^*, \cdots, x_{i-1}^*  , x_i^* , x_{i+1}^*, \cdots, x_{n}^*)
\nonumber \\
&  -  
r(x_1^*, \cdots, x_{i-1}^*  , 0 ,  x_{i+1}^*, \cdots, x_{n}^*)  \}
\quad \textrm{by Assumption~\ref{asp:decreasingMarginalReturn}} 
\nonumber \\
=
& \sum_{i=1}^n \{ r(\bm{x}^*) 
-  
r(0 , \bm{x}_{-i}^*)  \}.
\end{align}

Then, we can deduce:
{\allowdisplaybreaks
\begin{align*}
     & r(\bm{x}^* )
    -\sum_{i=1}^{n}c(x_i^*, \gamma_i)
    \nonumber \\
    \ge &
    \sum_{i=1}^{n} [ 
    r(\bm{x}^{*})
    -
    r((0,\bm{x}_{-i}^*))]
    - c(x_i^*,\gamma_i)\} 
    \qquad  \textrm{by Eq.~\eqref{eq:proofWBB2}}
    \nonumber \\
    = &
        \sum_{i=1}^{n} [ 
        r(\bm{x}^{*})
    -\sum_{k=1}^{n}c(x^{*}_k,\gamma_k)]
    \nonumber \\
    &
     - \sum_{i=1}^{n} [r((0, \bm{x}_{-i}^*) )
    - \sum_{k=1, \ne i}^{n}
        c(x^{*}_k,\gamma_k)]
    \nonumber \\
    = &
     \sum_{i=1}^{n} [r(\bm{x}^{*})
    -\sum_{k=1}^{n}c(x^{*}_k,\gamma_k)]
    \nonumber \\
    &
    - \sum_{i=1}^{n} [ r((0, \bm{x}_{-i}^*) )
    - c(0 ,\gamma_i)
    - \sum_{k=1, \ne i}^{n}
        c(x^{*}_k,\gamma_k)]
    \\
    &
    \qquad \qquad \qquad \qquad \qquad \qquad 
    \textrm{by Assumption~\ref{asp:zeroProcurement}}
    \nonumber \\
    = &
    \sum_{i=1}^{n}[
        S(\bm{x}^{*},\bm{\gamma})
        - S((0, \bm{x}_{-i}^*) ,\bm{\gamma})]
    \nonumber \\
    \ge &
    \sum_{i=1}^{n}[
        \mathcal{S}(\bar{\bm{x}},\bm{\gamma})
        - \mathcal{S}((0, \bar{\bm{x}}_{-i}) ,\bm{\gamma})]
    \\
    &    
    \qquad \qquad \qquad  \textrm{by definition of $\mathcal{S}(\cdot)$ and Eq.~\eqref{eq:proofWBB1}}.
        \nonumber
\end{align*}
}
From the above, it follows that:
\begin{align}
    r(\bm{x}^*) 
    & \ge
    \sum_{i=1}^{n}[
        \mathcal{S}(\bar{\bm{x}},\bm{\gamma},\bm{\theta})
        + c(x^{*}_i,\gamma_i)
        - \mathcal{S}((0, \bar{\bm{x}}_{-i}) ,\bm{\gamma}, \bm{\theta})]
        \nonumber\\
    & =
    \sum_{i=1}^{n}p_i(\bar{\bm{x}},\bm{\gamma}) 
    \qquad \qquad \qquad \textrm{by Eq.~\eqref{eq:proofWBB0}}.
    \nonumber
\end{align}
Consequently, the revenue of the coordinator is sufficient to cover the total payment to all suppliers.
\end{proof}

\subsection{Proof of Proposition~\ref{proposition:fitcurve}}
\label{appx:interpolation}

\subsubsection{Dominant-strategy incentive compatibility.}

The \pvcg{} payments derived from the interpolated curves also ensure dominate-strategy incentive compatibility. Specifically:
\begin{align}
    \label{eq:DICfitted}
       & 
       % \mathbb{E}_{\bar{\bm{x}}_{-i}, \bm{\gamma}_{-i}}
       [\hat{p}_i(
            (\bar{x}_i,\hat{{\bm{x}}}_{-i}),
            (\gamma_i, \hat{\bm{\gamma}}_{-i})
            )
         - c(
                \hat{x}^{*}_i(
                (\bar{x}_i,\hat{\bar{\bm{x}}}_{-i}),
            (\gamma_i, \hat{\bm{\gamma}}_{-i})
            ),
                \gamma_i
        )]
                \nonumber \\
        \ge &
        % \mathbb{E}_{\bar{\bm{x}}_{-i}, \bm{\gamma}_{-i}} 
        [
        \hat{p}_i(
            (\hat{x}_i,\hat{\bm{x}}_{-i}),
            (\hat{\gamma}_i, \hat{\bm{\gamma}}_{-i})
            )
         - c(
                \hat{x}^{*}_i(
                (\hat{x}_i,\hat{\bm{x}}_{-i}),
            (\hat{\gamma}_i, \hat{\bm{\gamma}}_{-i})
            ),
                \gamma_i
        )],
        \nonumber \\
        & 
        \quad \qquad\qquad\qquad\qquad 
        \forall i\in N, \, \bar{x}_i,\, \gamma_i, \, \hat{x}_i ,\, \hat{\gamma}_i,\,
        \hat{\bm{x}}_{-i}, \hat{\bm{\gamma}}_{-i}.
\end{align}
Here, $\hat{\bm{x}}^*$ and $\hat{\bm{p}}$ are determined from the interpolated curves.

\begin{proof}
Drawing a parallel to the proof in Appendix~\ref{appx:proofIC}, since supplier $i$ will avoid reporting a capacity limit that would result in an optimal allocation surpassing its true capacity limit, we have:
\[
\hat{x}_i^*( (\hat{x}_i, \hat{\bm{x}}_{-i}), (\hat{\gamma}_i, \hat{\bm{\gamma}}_{-i})) \le \bar{x}_i,
\]

Using a logic analogous to Eq.~\eqref{eq:proofDIC1}, we can  deduce:
\begin{align}
    \label{eq:proofDICfitted1}
    & \hat{\mathcal{S}}( (\bar{x}_i, \hat{\bm{x}}_{-i}), (\gamma_i, \hat{\bm{\gamma}}_{-i}))
    \nonumber \\
    \ge & 
    \hat{\mathcal{S}}( (\hat{x}_i, \hat{\bm{x}}_{-i}), (\hat{\gamma}_i, \hat{\bm{\gamma}}_{-i})) 
    + \hat{c}(\hat{x}_i^*( (\hat{x}_i, \hat{\bm{x}}_{-i}), (\hat{\gamma}_i, \hat{\bm{\gamma}}_{-i})), \hat{\gamma}_i) \nonumber \\
    &  - \hat{c}(\hat{x}_i^*( (\hat{x}_i, \hat{\bm{x}}_{-i}), (\hat{\gamma}_i, \hat{\bm{\gamma}}_{-i})), \gamma_i)
    .
\end{align} 
In the above, $\hat{c}(\cdot, \hat{\gamma}_i) $ and  $\hat{c}(\cdot, \gamma_i) $ denote the interpolated cost curve for the reported cost type $\hat{\gamma}_i$ and the true cost type $\gamma_i$, respectively.  The premises of Proposition~\ref{proposition:fitcurve} ensures:
\begin{align*}
   & c(
                \hat{x}^{*}_i(
                (\bar{x}_i,\hat{\bm{x}}_{-i}),
            (\gamma_i, \hat{\bm{\gamma}}_{-i})
            ),
                \gamma_i
        )
\\
\equiv & \hat{c}(
                \hat{x}^{*}_i(
                (\bar{x}_i,\hat{\bm{x}}_{-i}),
            (\gamma_i, \hat{\bm{\gamma}}_{-i})
            ) , \gamma_i) ,   
\\
&
c(\hat{x}_i^*( (\hat{x}_i, \hat{\bm{x}}_{-i}), (\hat{\gamma}_i, \hat{\bm{\gamma}}_{-i})), \gamma_i) 
\\
\ge &
\hat{c}(\hat{x}_i^*( (\hat{x}_i, \hat{\bm{x}}_{-i}), (\hat{\gamma}_i, \hat{\bm{\gamma}}_{-i})), \gamma_i).
\end{align*}

From the above, we obtain:
\begin{align}
\label{eq:proofDICfit2}
& \textrm{LHS} - \textrm{RHS of Eq.~\eqref{eq:DICfitted}}
\nonumber\\
\ge &  
% \mathbb{E}_{\bar{\bm{x}}_{-i}, \bm{\gamma}_{-i}} [
\textrm{LHS} - \textrm{RHS of Eq.~\eqref{eq:proofDICfitted1}}
% + \mathbb{E}_{\bar{\bm{x}}_{-i}, \bm{\gamma}_{-i}}  
\nonumber \\
& 
+ [c(\hat{x}_i^*( (\hat{x}_i, \hat{\bm{x}}_{-i}), (\hat{\gamma}_i, \hat{\bm{\gamma}}_{-i})), \gamma_i) 
\nonumber \\
& \quad - \hat{c}(\hat{x}_i^*( (\hat{x}_i, \hat{\bm{x}}_{-i}), (\hat{\gamma}_i, \hat{\bm{\gamma}}_{-i})), \gamma_i)
]
\nonumber \\
& -   [ 
c(
                \hat{x}^{*}_i(
                (\bar{x}_i,\hat{\bm{x}}_{-i}),
            (\gamma_i, \hat{\bm{\gamma}}_{-i})
            ),
                \gamma_i
        )
\nonumber \\ 
& \quad - \hat{c}(
                \hat{x}^{*}_i(
                (\bar{x}_i,\hat{\bm{x}}_{-i}),
            (\gamma_i, \hat{\bm{\gamma}}_{-i})
            ) , \gamma_i) ]
\ge  0. 
\end{align}

% The premise that the fitted curves, based on the reported parameters, yield the same optimal allocation as the real curves implies:
% \[
% \hat{x}_i^*( (\hat{x}_i, \bar{\bm{x}}_{-i}), (\hat{\gamma}_i, \bm{\gamma}_{-i}))
% = x_i^*( (\hat{x}_i, \bar{\bm{x}}_{-i}), (\hat{\gamma}_i, \bm{\gamma}_{-i})) .
% \]

% Since we assume the fitted cost curve is tangent to the real cost curve at point $\hat{x}_i^*$ and lies below it elsewhere, it can be deduced that the function:
% \[
%     c(x_i, \gamma_i) - \hat{c}(x_i, \gamma_i),
% \]
% attains its lowest value, 0,  at $\hat{x}_i^*(  (\hat{x}_i,\bar{\bm{x}}_{-i}),
%             (\hat{\gamma}_i, \bm{\gamma}_{-i}))$. 
            
% Consequently, the second expectation term in the RHS of  Eq.~\eqref{eq:proofDICfit2} is greater than $0$, since it represents the discrepancy between an average centered at an arbitrary point and one centered at the minimum point. This leads to the conclusion:
% \[
%     \textrm{RHS of Eq.~\eqref{eq:proofDICfit2} } \ge 0 .
% \]

As a result, the inequality in  Eq.~\eqref{eq:DICfitted} is valid.

\end{proof}

\subsubsection{Pareto efficiency, individual rationality, and weak budget balance}

\begin{proof}

By the premises of Proposition~\ref{proposition:fitcurve}, the optimal allocation on the interpolated curves is consistent with the real curves.  Hence, the Pareto efficiency of \pvcg{}  on real curves guarantees the Pareto efficiency of the \pvcg{} sharing rule from interpolated curves.

Consider the ex-post utility of supplier $i$ as follows:
\begin{align}
    % \label{equation:IRproof}
    & \hat{p}_i(\bar{\bm{x}},\bm{\gamma})
    - c(\hat{x}_{i}^{*}(\bar{\bm{x}},\bm{\gamma}),\gamma_i)
    \nonumber \\
    = &
    \hat{\mathcal{S}}(\bar{\bm{x}},\bm{\gamma}) 
        - \hat{\mathcal{S}}((0,\bar{\bm{x}}_{-i}),\bm{\gamma})
        + \hat{c}(\hat{x}_{i}^{*}(\bar{\bm{x}},\bm{\gamma}),\gamma_i)
        - c(\hat{x}_{i}^{*}(\bar{\bm{x}},\bm{\gamma}),\gamma_i)
    \nonumber \\
    = &
    \hat{\mathcal{S}}(\bar{\bm{x}},\bm{\gamma}) 
        - \hat{\mathcal{S}}((0,\bar{\bm{x}}_{-i}),\bm{\gamma})  
    \quad   \textrm{by overlap of} \, \hat{c}(\cdot) \, \textrm{and}\, c(\cdot) \, \textrm{at} \, \hat{x}_i^*
    \nonumber \\
    \ge &
    0 \qquad \qquad \qquad \qquad \qquad \qquad \quad \,\,
        \textrm{by definition of }    \hat{\mathcal{S}}(\bar{\bm{x}},\bm{\gamma}). 
    \nonumber
\end{align}
Hence, the \pvcg{} sharing rule derived from interpolated curves retains the property of individual rationality.

Assuming the interpolated revenue curve $\hat{r}(\cdot)$ also adheres to Assumption ~\ref{asp:decreasingMarginalReturn}, and drawing parallels with the  proof in  Appendix~\ref{appx:proofWBB}, we can establish:
\[
    \hat{r}(\hat{\bm{x}}^*(\bar{\bm{x}}, \bm{\gamma}) ) \ge \sum_{i=1}^n \hat{p}_i(\bar{\bm{x}}, \bm{\gamma}).
\]

Given the alignment of the interpolated revenue curve with the real revenue curve at 
 $\hat{\bm{x}}^*(\bar{\bm{x}}, \bm{\gamma})$, we obtain:
\[
    r(\hat{\bm{x}}^*(\bar{\bm{x}}, \bm{\gamma})) = \hat{r}(\hat{\bm{x}}^*(\bar{\bm{x}}, \bm{\gamma})) \ge \sum_{i=1}^n \hat{p}_i(\bar{\bm{x}}, \bm{\gamma}).
\]
Thus, the \pvcg{} sharing rule derived from interpolated curves upholds the property of weak budget balance.

\end{proof}

\begin{algorithm}[b!]
    \caption{Protocol \pvcg{}}
    \label{alg:pvcg}
    \KwIn{Coordinator's input: Noisy revenue oracle; \\
          $\qquad \quad$ Suppliers' private input: Noisy cost oracles; }
    \KwOut{\pvcg{} payments: $\bm{p} \in \mathbb{R}^n$, \\
           $\qquad $ Optimal procurement level: $\bm{x}^* \in \mathbb{R}^n$, \\
           $\qquad $ Optimal allocation excluding supplier $i$: $\{\bm{z}_i^* \in \mathbb{R}^n \}_{i \in N}$;}
   Suppliers report their capacity limits and marginal costs at the quantiles of these capacity limits\;
    $\bm{x} \leftarrow \hat{\bm{x}}/4$, $\hat{\bm{x}}^* \leftarrow \bm{0}$\;
    \For{  current\_epoch \textbf{\texttt{in}} $1$ \texttt{to} MAX\_EPOCHS}{
         \tcc{ *************** R-Step ****************} 
        Suppliers report their marginal costs at the quantiles of the current procurement levels\;
        The coordinator measures revenue gradients at the quantiles of the current procurement level\;
        \tcc{ *************** I-Step ****************} 
        Interpolate cost curves from historical marginal costs\;
         Interpolate the revenue curve from historical gradients\;
        \tcc{ *************** M-Step ****************} 
        $\bm{x} \leftarrow$ Update using one step of gradient descent based on current marginal costs and the revenue gradient\;
        $\hat{\bm{x}}^* \leftarrow$ Determine the optimal procurement level using interpolated curves\;
        Update $\bm{x}$ toward $\hat{\bm{x}}^*$ by the rapid adjustment coefficient\;
    }
    $\bm{x}^* \leftarrow \bm{x}$\;
   $\bm{z}_i^* \leftarrow$ Compute the optimal procurement level excluding supplier $i$ using interpolated curves, for each $i \in N$  \;
    $\bm{p} \leftarrow $ \pvcg{} payments using Eq.~\eqref{eq:pvcgFromDerivatives}\;
    \Return $\bm{p}, \bm{x}^*, \{\bm{z}_i^*\}_{i \in N}$
\end{algorithm}

% \vspace{-0.5cm}

\section{Experimental Details}
\label{appx:experiment}

The pseudo-code for our experiment is presented in Algorithm~\ref{alg:pvcg}. 
All computations were performed on a machine equipped with an Intel i7 processor and 64GB of RAM.

Parameters used in our experiments are outlined in Table~\ref{tab:pars}. These values were chosen to ensure that half of the suppliers reach their capacity limits at the optimal procurement level. 
To prevent over-fitting at a local minimum, we allowed the agents to measure the marginal costs and revenue gradients at the quantiles corresponding to their current procurement levels during each interactive epoch.
In the gradient descent optimization, we included a momentum term with a coefficient of $0.9$ and allowed the allocation to update rapidly toward the interpolated allocation using a coefficient of $0.1$. We employed different learning rates: one for the gradient descent during interactive epochs, which incurs higher costs due to real-world interactions, and another for the local gradient descent on interpolated curves. For the spline regression, we utilized the cubic B-spline model and set the degree of freedom (DF) to $6$. We also removed outliers from the regressions to obtain more robust estimations.

\begin{table}[h]
  \caption{Experiment Parameters}
  \label{tab:pars}
  \begin{tabular}{cc}
    \toprule
    Parameter& Value\\
    \midrule
    Number of suppliers, $n$ & $10$ \\
    Capacity limits, $\bar{\bm{x}}$ & $[1.,2.,3.,4.,5.,6.,7.,8.,9.,10.]$ \\
    Cost coefficients, $\bm{\kappa}$ & $[1.,2.,3.,4.,5.,6.,7.,8.,9.,10.]$\\
    Revenue coeffcient, $\rho$ & $500.$\\ 
    Noise level & $10\%$ \\
    Interactive learning rate & $0.1$ \\
    Max no. of interactive epoches & $50$ \\
    Local learning rate & $0.001$ \\
    Max no. of local epoches & $10,000$ \\
    Momentum coefficient & $0.9$ \\
    Rapid adjustment coefficient & $0.1$ \\
    DF for cubic B-spline & $6$ \\
  \bottomrule
\end{tabular}
\end{table}

\end{document}